\pdfoutput=1

\documentclass[final,10pt]{elsarticle}
\makeatletter
\def\ps@pprintTitle{%
 \let\@oddhead\@empty
 \let\@evenhead\@empty
 \def\@oddfoot{}%
 \let\@evenfoot\@oddfoot}
\makeatother
\usepackage{graphicx}
\usepackage{rotating}
\usepackage{epstopdf, epsfig}
\usepackage[utf8]{inputenc} 
\usepackage[T1]{fontenc}    
\usepackage{hyperref}       
\usepackage{url}            
\usepackage{amsfonts}       
\usepackage{amsthm}       
\usepackage{mathtools}       
\usepackage{nicefrac}       
\usepackage{microtype}      
\usepackage{graphics}
\usepackage{pgfplots}
\usepackage{pgfplotstable}
\usepgfplotslibrary{fillbetween}
\usepackage{tikz}
\usepackage{bm}
\usepackage{xfrac}
\usepackage{float}
\usepackage{color}
\usepackage{algorithm}
\usepackage[noend]{algpseudocode}
\usepackage{mathtools}
\usepackage{amssymb}
\usepackage{amsmath}
\usepackage{standalone}
\usepackage{fullpage}
\usepackage[keeplastbox]{flushend}
\usepackage{comment}

\definecolor{green}{RGB}{0,102,0}

\usepackage{bm}
\usepackage{amssymb}
\usepackage{mathtools}
\usepackage{nicefrac}
\usepackage{subfig}

\usepackage{amsmath}
\usepackage{amsfonts}
\usepackage{bm}
\usepackage{subfloat}
\usepackage{mathtools}
\usepackage{soul}
\usepackage{xcolor}
\usepackage{comment}

\newcommand{\tran}{^\mathsf{T}}
\newcommand{\fro}{_\mathsf{F}}

\newcommand{\dofFOM}{N}
\newcommand{\dofROM}{p}
\newcommand{\nParams}{n_{\paramSymb}}
\newcommand{\nSteps}{n_{\timeSymb}}
\newcommand{\nTrain}{n_{\text{train}}}

\newcommand{\dofVars}{n_{\solutionSymb}}
\newcommand{\dofFOMx}{n_1}
\newcommand{\dofFOMy}{n_2}

\newcommand{\primSolutionSymb}{w}
\newcommand{\primSolution}{\bm{w}}

\newcommand{\diffusionCoeff}{\nu}

\newcommand{\sourceTermSymb}{q}
\newcommand{\sourceTerm}{\bm{\sourceTermSymb}}

\newcommand{\temperatureSymb}{T}
\newcommand{\temperature}{\mathrm{\temperatureSymb}}
\newcommand{\Htwo}{\mathrm{H}_2}
\newcommand{\Otwo}{\mathrm{O}_2}
\newcommand{\HtwoO}{\mathrm{H}_2\mathrm{O}}

\newcommand{\solutionSymb}{u}
\newcommand{\solution}{\bm{\solutionSymb}}
\newcommand{\reducedSolution}{\hat{\bm{\solutionSymb}}}
\newcommand{\approxSolution}{\tilde{\bm{\solutionSymb}}}

\newcommand{\timeSymb}{t}
\newcommand{\finalTime}{T}
\newcommand{\paramSpaceSymb}{\mathcal D}
\newcommand{\paramSpaceTrain}{\paramSpaceSymb_{\text{train}}}
\newcommand{\paramSpaceVal}{\paramSpaceSymb_{\text{val}}}
\newcommand{\paramSpaceTest}{\paramSpaceSymb_{\text{test}}}

\newcommand{\paramSymb}{\mu}
\newcommand{\PDEparams}{\bm{\paramSymb}}
\newcommand{\PDEparamsTrain}{\PDEparams_{\text{train}}}

\newcommand{\PDEparamsTest}{\PDEparams_{\text{test}}}

\newcommand{\positionalSymb}{x}
\newcommand{\positions}{\bm{\positionalSymb}}

\newcommand{\initCond}{\solution^{0}(\PDEparams)}
\newcommand{\initCondReduced}{\reducedSolution^{0}(\PDEparams)}

\newcommand{\velocitySymb}{f}
\newcommand{\velocity}{\bm{\velocitySymb}}
\newcommand{\reducedVelocity}{\hat{\bm{\velocitySymb}}}
\newcommand{\nodeVelocity}{\velocity_{\NODEParams}}
\newcommand{\pnodeVelocity}{\reducedVelocity_{\NODEParams}}

\newcommand{\NODEParams}{\Theta}
\newcommand{\NNParams}{\pmb{\theta}}

\newcommand{\NNSymb}{\pmb{h}}
\newcommand{\encoder}{\NNSymb_{\text{enc}}}
\newcommand{\decoder}{\NNSymb_{\text{dec}}}

\newcommand{\nonlinearMapDecoder}{\pmb{d}}

\newcommand{\encParams}{\NNParams_{\text{enc}}}
\newcommand{\decParams}{\NNParams_{\text{dec}}}

\newcommand{\lossSymb}{L}

\newcommand{\solutionCollectionSymbol}{U}
\newcommand{\solutionMatrix}{\pmb{\solutionCollectionSymbol}}
\newcommand{\solutionTensor}{\pmb{\mathcal \solutionCollectionSymbol}}

\newcommand{\approxSolutionMatrix}{\tilde{\pmb{\solutionCollectionSymbol}}}

\newcommand{\kernelLength}{\kappa}
\newcommand{\nKernels}{n_{\kernelLength}}
\newcommand{\strides}{s}

\newcommand{\dimSymb}{d}
\newcommand{\inputDim}{\dimSymb_{\text{in}}}
\newcommand{\outputDim}{\dimSymb_{\text{out}}}

\newcommand{\forcingSymb}{f}
\newcommand{\forcing}{\bm{\forcingSymb}}

\newcommand{\source}{\bm{g}}

\newcommand{\densitySymb}{\rho}
\newcommand{\VelocitySymb}{u}
\newcommand{\energySymb}{e}
\newcommand{\potentialEnergy}{\epsilon}
\newcommand{\pressureSymb}{p}
\newcommand{\areaSymb}{A}
\newcommand{\heatRatio}{\gamma}
\newcommand{\gasConstant}{R}
\newcommand{\totalTemp}{\temperatureSymb_\text{total}}
\newcommand{\totalPressure}{\pressureSymb_{\text{total}}}
\newcommand{\MachSymb}{M}
\newcommand{\midArea}{_{m}}
\newcommand{\soundSymb}{c}

\newcommand{\hiddenStateSymb}{z}
\newcommand{\hiddenState}{\bm{\hiddenStateSymb}}

\newcommand{\timeDeriv}[1]{\dot{#1}}

\newcommand{\RR}[1]{\mathbb{R}^{#1}}

\theoremstyle{definition}

\begin{document}
\numberwithin{equation}{section}
\begin{frontmatter}
	\title{Parameterized Neural Ordinary Differential Equations: \\Applications to Computational Physics Problems}

\author[sandia]{Kookjin Lee\corref{sandiacor}}
\ead{koolee@sandia.gov}
\author[sandia]{Eric J. Parish}
\ead{ejparis@sandia.gov}

\address[sandia]{Sandia National Laboratories}
\cortext[sandiacor]{7011 East Ave, MS 9159, Livermore, CA 94550.}

\begin{abstract}
This work proposes an extension of neural ordinary differential equations (NODEs) by introducing an additional set of ODE input parameters to NODEs. This extension allows NODEs to learn multiple dynamics specified by the input parameter instances. Our extension is inspired by the concept of parameterized ordinary differential equations, which are widely investigated in computational science and engineering contexts, where characteristics of the governing equations vary over the input parameters. We apply the proposed parameterized NODEs (PNODEs) for learning latent dynamics of complex dynamical processes that arise in computational physics, which is an essential component for enabling rapid numerical simulations for time-critical physics applications. For this, we propose an encoder-decoder-type framework, which models latent dynamics as PNODEs. We demonstrate the effectiveness of PNODEs with important benchmark problems from computational physics.
\end{abstract}

\end{frontmatter}

\section{Introduction}
Numerical simulations of dynamical systems described by systems of ordinary differential equations (ODEs)\footnote{Such systems of ODEs often arise from spatial discretization of time-dependent partial differential equations.} play essential roles in various engineering and applied science applications. Such examples include predicting input/output responses, design, and optimization \cite{quarteroni2015reduced}.
These ODEs and their solutions often depend on a set of \textit{input parameters}, and such ODEs are denoted as \textit{parameterized ODEs}. Examples of such input parameters within the context of fluid dynamics include Reynolds number and Mach number. In many important scenarios, \textit{high-fidelity} solutions of parameterized ODEs are required to be computed i) for many different input parameter instances (i.e., \textit{many-query} scenario) or ii) in \textit{real time} on a new input parameter instance. A single run of a high-fidelity simulation, however, often requires fine spatiotemporal resolutions. Consequently, performing real-time or multiple runs of a high-fidelity simulation can be computationally prohibitive.

To mitigate this computational burden, many model-order reduction approaches have been proposed to replace costly high-fidelity simulations. The common goal of these approaches is to build a reduced-dynamical model with lower complexity than that of the high-fidelity model, and to use the reduced model to compute approximate solutions for any new input parameter instance. In general, model-order reduction approaches consist of two components: i) a low-dimensional latent-dynamics model, where the computational complexity is very low, and ii) a (non)linear mapping that constructs high-dimensional approximate states (i.e., solutions) from the low-dimensional states obtained from the latent-dynamics model. In many studies, such models are constructed via \textit{data-driven} techniques in the following steps: i) collect solutions of high-fidelity simulations for a set of training parameter instances, ii) build a parameterized surrogate model, 
and iii) fit the model by training with the data collected from the step i).

In the field of deep-learning, similar efforts have been made for learning latent dynamics of various physical processes \cite{karl2016deep,chen2018neural,morton2018deep,rubanova2019latent,fulton2019latent}. 
Neural ordinary differential equations (NODEs), a method of learning time-continuous dynamics in the form of a system of ordinary differential equations from data, comprise a particularly promising approach for learning latent dynamics of dynamical systems. NODEs have been studied in \cite{weinan2017proposal,chen2018neural,haber2017stable,ruthotto2019deep,lu2018beyond,ciccone2018nais,gholami2019anode}, and this body of work has demonstrated their ability to successfully learn latent dynamics and to be applied to downstream tasks \cite{chen2018neural,rubanova2019latent}.

Because NODEs learn latent dynamics in the form of ODEs, NODEs have a naturally good fit as a latent-dynamics model in reduced-order modeling of physical processes and have been applied to several computational physics problems including turbulence modeling \cite{portwood2019turbulence,maulik2020time} and future states predictions in fluids problems \cite{ayed2019learning}. As pointed out in \cite{dupont2019augmented,chalvidal2020neural}, however, NODEs learn a single set of network weights, which fits best for a given training data set. This results in an NODE model with limited expressibility and often leads to unnecessarily complex dynamics \cite{finlay2020train}. To overcome this shortcoming, we propose to extend NODEs to have a set of input parameters that specify the dynamics of the NODE model, which leads to parameterized NODEs (PNODEs). With this simple extension, PNODEs can represent multiple trajectories such that the dynamics of each trajectory are characterized by the input parameter instance. 

The main contributions of this paper are 
\begin{itemize}
\item an extension to NODEs that enables them to learn multiple trajectories with a single set of network weights; even for the same initial condition, the dynamics can be different for different input parameter instances,
\item a framework for learning latent dynamics of parameterized ODEs arising in computational physics problems,
\item a demonstration of the effectiveness of the proposed framework with advection-dominated benchmark problems, which are a class of problems where classical linear latent-dynamics learning methods (e.g., principal component analysis) often fail to learn accurately~\cite{lee2019deep}.
\end{itemize}

\section{Related work}
\paragraph{Classical reduced-order modeling}
Classical reduced-order modeling (ROM) techniques rely heavily on linear methods such as the proper orthogonal decomposition (POD) \cite{holmes2012turbulence}, which is analogous to principal component analysis \cite{hotelling1933analysis}, for constructing the mappings between a high-dimensional space and a low-dimensional space. These ROMs then identify the latent-dynamics model by executing a (linear) projection process on the high-dimensional equations e.g., Galerkin projection \cite{holmes2012turbulence} or least-square Petrov--Galerkin projection \cite{carlberg2013gnat,carlberg2017galerkin}. We refer readers to  \cite{benner2015survey,benner2017model,ohlberger2015reduced} for a complete survey on classical methods.

\paragraph{Physics-aware deep-learning-based reduced-order modeling } 
Recent work has extended classical ROMs by replacing proper orthogonal decomposition with nonlinear dimension reduction techniques emerging from deep learning~\cite{kashima2016nonlinear,hartman2017deep,fulton2019latent,lee2019deep,lee2020model,kim2020fast}. These approaches operate by identifying a nonlinear mapping (via, e.g., convolutional autoencoders) and subsequently identifying the latent dynamics as certain residual minimization problems~\cite{fulton2019latent,lee2019deep,lee2020model,kim2020fast}, which are defined on the latent space and are derived from the governing equations. In \cite{kashima2016nonlinear,hartman2017deep}, the latent dynamics is identified by simply projecting the governing equation using the encoder, which may leads to kinematically inconsistent dynamics.

Another class of physics-aware methods include explicitly modeling time integration schemes \cite{pawar2019deep,san2019artificial,xie2019non,geneva2020modeling}, adaptive basis selection \cite{rim2020depth}, and adding stability/structure-preserving constraints in the latent dynamics \cite{erichson2019physics,hernandez2020deep}. We emphasize that our approach is closely related to \cite{san2019artificial}, where neural networks are trained to approximate the action of the first-order time-integration scheme applied to latent dynamics and, at each time step, the neural networks take a set of problem-specific parameters as well as reduced state as an input. Thus, our approach can be seen as a time-continuous generalization of the approach in \cite{san2019artificial}.

\paragraph{Purely data-driven deep-learning-based reduced-order modeling}
Another approach for developing deep-learning-based ROMs is to learn  
both nonlinear mappings and latent dynamics in purely data-driven ways. Latent dynamics are modeled as recurrent neural networks with long short-term memory (LSTM) units along with linear POD mappings \cite{wang2018model,rahman2019nonintrusive,maulik2020time} or nonlinear mappings constructed via (convolutional) autoencoders \cite{gonzalez2018deep,wiewel2019latent,maulik2020reduced,tencer2020enabling}. In \cite{portwood2019turbulence,maulik2020time}, latent dynamics and nonlinear mappings are modeled as neural ODEs and autoencoders, respectively;  
in \cite{otto2017linearly,lusch2017deep,takeishi2017learning,morton2018deep}, autoencoders are used to learn approximate invariant subspaces of the Koopman operator. Relatedly, there have been studies on learning direct mappings via e.g., a neural network, from problem-specific parameters to either latent states or approximate solution states \cite{swischuk2019projection,fresca2020comprehensive,renganathan2020machine,chenaphysics,kast2020non,wang2019non}, where the latent states are computed by using autoencoder or linear POD.

\paragraph{Enhancing NODE} Augmented NODEs \cite{dupont2019augmented} extends NODEs by augmenting additional state variables to hidden state variables, which allows NODEs to learn dynamics using the additional dimensions and, consequently, to have increased expressibility. ANODE \cite{gholami2019anode} discretize the integration range into a fixed number of steps (i.e., checkpoints) to mitigate numerical instability in the backward pass of NODEs; ACA~\cite{zhuang2020adaptive} further extends this approach by adopting adaptive stepsize solver in the bardward pass. ANODEV2 \cite{zhang2019anodev2} proposes a coupled system of neural ODEs, where both hidden state variables and network weights are allowed to evolve over time and their dynamics are approximated as neural networks. Neural optimal control \cite{chalvidal2020neural} formulates an NODE model as a controlled dynamical system and infers optimal control via an encoder network 
This formulation results in an NODE that adjusts the dynamics for different input data. Moreover, improved training strategies for NODEs have been studied in \cite{finlay2020train} and an extension of using spectral elements in discretizations of NODE has been proposed in \cite{quaglino2019snode}.

\section{Neural ODE}
Neural ODEs (NODEs) are a family of deep neural network models that parameterize the time-continuous dynamics of hidden states using a system of ODEs:
\begin{equation}\label{eq:node}
\frac{d \hiddenState(\timeSymb)}{d\timeSymb} = \nodeVelocity(\hiddenState(\timeSymb),\timeSymb;\Theta),
\end{equation} 
where $\hiddenState(\timeSymb)$ is a time-continuous representation of a hidden state, $\nodeVelocity$ is a parameterized velocity function, which defines the dynamics of hidden states over time, and $\Theta$ is a set of neural network weights. Given the initial condition $\hiddenState(0)$ (i.e., input), a hidden state at any time index $\hiddenState(\timeSymb)$ can be obtained by solving the initial value problem (IVP)~\eqref{eq:node}. To solve the IVP, a black-box differential equation solver can be employed and the hidden states can be computed with the desired accuracy:
\begin{equation}
\hiddenState^{1},\ldots,\hiddenState^{\nSteps} = \text{ODESolve}(\hiddenState(0),\nodeVelocity,\timeSymb_1,\ldots,\timeSymb_{\nSteps}).
\end{equation}
In the backward pass, as proposed in \cite{chen2018neural}, gradients are computed by solving another system of ODEs, which are derived using the adjoint sensitivity method \cite{pontryagin1962mathematical}
, which allows memory efficient training of the NODE model. As pointed out in the papers \cite{dupont2019augmented,chalvidal2020neural}, an NODE model learns a single dynamics for the entire data distribution and, thus, results in a model with limited expressivity. 

\section{Parameterized neural ODE}
To resolve this, we propose a simple, but powerful extension of neural ODE. We refer to this extension a ``parameterized neural ODEs'' (PNODEs): 
\begin{equation}
\frac{d \hiddenState(\timeSymb;\PDEparams)}{d\timeSymb} = \nodeVelocity(\hiddenState(\timeSymb;\PDEparams),\timeSymb;\PDEparams, \Theta),
\end{equation} 
with a parameterized initial condition $\hiddenState^{0}(\PDEparams)$, where $\PDEparams=[\paramSymb_1,\ldots,\paramSymb_{\nParams}] \in \paramSpaceSymb \subset \RR{\nParams}$ denotes problem-specific input parameters. Inspired by the concept of ``parameterized ODEs'', where the ODEs depend on the input parameters, this simple extension allows NODE to have multiple latent trajectories that depend on the input parameters. This extension only requires minimal modifications in the definition of the velocity function $\nodeVelocity$ and can be trained/deployed by utilizing the same mathematical machinery developed for NODEs in the forward pass (i.e., via a black-box ODE solver) and the backward pass (i.e., via the adjoint-sensitivity method). In practice, $\nodeVelocity$ is approximated as a neural network which takes $\hiddenState$ and $\PDEparams$ as an input and then produces $\frac{d \hiddenState}{d\timeSymb}$ as an output. 


\section{Applications to computational physics problems}
{We now investigate PNODEs within the context of performing model reduction of computational physics problems. We start by formally introducing the full-order model that we seek to reduce. We then describe our proposed framework, which uses PNODEs (or NODEs) as the reduced-order (latent-dynamics) model.} 
\subsection{Full-order model (FOM)}
The full-order model (FOM) corresponds to a parameterized system of ordinary differential equations (ODEs):
\begin{equation}\label{eq:fom}
\timeDeriv{\solution} = \velocity(\solution,\timeSymb;\PDEparams), \quad \solution(0;\PDEparams) = \initCond,
\end{equation}
where $\solution(\timeSymb;\PDEparams)$, $\solution: [0,\finalTime]\times\paramSpaceSymb \rightarrow \RR{\dofFOM}$ denotes the state, which is implicitly defined as the solution to the system of ODEs. Here, $\PDEparams \in \paramSpaceSymb$ denotes the ODE parameters that characterize physical properties (e.g., boundary conditions, forcing terms), $\paramSpaceSymb \subset \RR{\nParams}$ denotes the parameter space, where $\nParams$ is the number of parameters, and $\finalTime$ denotes the final time. The initial state is specified by the parameterized initial condition $\initCond$, $\solution^{0}: \paramSpaceSymb \rightarrow \RR{\dofFOM}$. Lastly, $\velocity(\solution,\timeSymb;\PDEparams)$, $\velocity: \RR{\dofFOM}\times [0,\finalTime]\times\paramSpaceSymb \rightarrow \RR{\dofFOM}$ denotes the velocity and $\timeDeriv{\solution}$ denotes the differentiation of $\solution$ with respect to time $\timeSymb$. Solving \eqref{eq:fom} requires application of an ODE solver, where the computational complexity rapidly grows with degrees of freedom $\dofFOM$ (e.g., $\dofFOM\sim 10^{7}$ {for many practical problems in computational physics}).

\subsection{Reduced-order model (ROM)}
Reduced-order modeling mitigates the high cost associated with solving the FOM by operating on reduced computational models that comprise i) a (non)linear mapping that constructs high-dimensional states from reduced states and ii) a low-dimensional latent-dynamics model for the reduced states. We denote the mapping from the reduced states to the high-dimensional states as $\nonlinearMapDecoder: \RR{p} \rightarrow \RR{N}$, and denote the latent dynamics model as the system of parameterized ODEs:
\begin{equation}\label{eq:ROM_eq}
\timeDeriv{\reducedSolution} = \reducedVelocity(\reducedSolution,\timeSymb;\PDEparams) , \quad \reducedSolution(0;\PDEparams) = \initCondReduced,
\end{equation}
where $\reducedSolution(\timeSymb;\PDEparams)$, $\reducedSolution: [0,\finalTime]\times\paramSpaceSymb \rightarrow \RR{\dofROM}$ denotes the reduced state, which is a low-dimensional representative state of the high-dimensional state (i.e., $\dofROM \ll \dofFOM$). Analogously, $\initCondReduced$, $\reducedSolution^{0}: \paramSpaceSymb \rightarrow \RR{\dofROM}$ denotes the reduced parameterized initial condition, and $\reducedVelocity(\reducedSolution,\timeSymb;\PDEparams)$, $\reducedVelocity: \RR{\dofROM}\times [0,\finalTime]\times\paramSpaceSymb \rightarrow \RR{\dofROM}$ denotes the reduced velocity. The objective of the ROM is to learn both a nonlinear mapping and a latent-dynamics model such that the ROM generates accurate approximate solutions to the full-order model solution, i.e., $\nonlinearMapDecoder(\reducedSolution) \approx \solution$. 
\subsection{Learning latent-dynamics with PNODE}
The aim here is to learn the latent dynamics with the PNODE: find a set of NODE parameters $\NODEParams$ such that
\begin{align*}
\timeDeriv{\reducedSolution} &= \pnodeVelocity(\reducedSolution,\timeSymb;\PDEparams, \NODEParams), 
\end{align*}
where $\pnodeVelocity(\cdot,\cdot;\cdot,\NODEParams): \RR{\dofROM}\times [0,\finalTime]\times\paramSpaceSymb \rightarrow \RR{\dofROM}$ 
denotes the reduced velocity, i.e., modeling a ROM (Eq.~\eqref{eq:ROM_eq})  as PNODE. 
To achieve this goal, we propose a framework, where
, besides a latent-dynamics model described by the PNODE, two additional functions are required:  i) an encoder, which maps a high-dimensional initial state $\initCond$ to a reduced initial state $\initCondReduced$, and ii) a decoder, which maps a set of reduced states $\reducedSolution^{k}, k=1,\ldots, \nSteps$ to a set of high-dimensional approximate states $\approxSolution^{k}, k=1,\ldots,\nSteps$. We approximate these functions with two neural networks: the encoder $\reducedSolution = \encoder(\solution;\encParams)$, $\encoder: \RR{\dofFOM} \rightarrow \RR{\dofROM}$ and the decoder $\approxSolution = \decoder(\reducedSolution;\decParams)$, $\decoder: \RR{\dofROM} \rightarrow \RR{\dofFOM}$ (i.e., $\nonlinearMapDecoder=\decoder$). 
Here, $\NNParams = (\encParams, \decParams)$ are the network weights.

\begin{figure}[t]
    \centering
    {\includegraphics[scale=1]{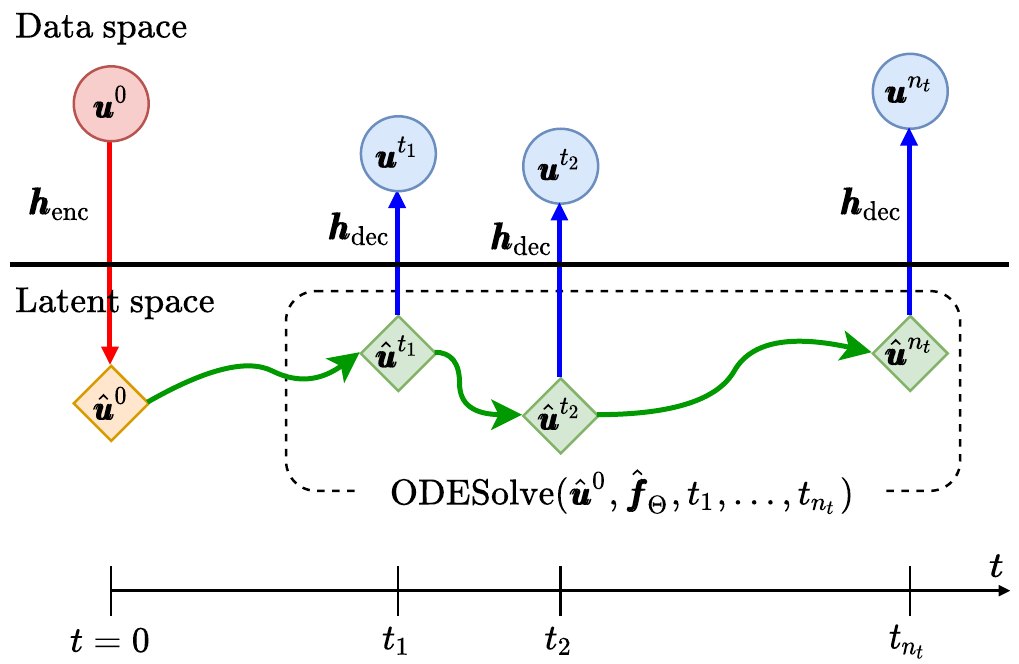} }
    \caption{The forward pass of the proposed framework: i) the encoder (red arrow), which provides a reduced initial state to the PNODE, ii) solving PNODE (or NODE) with the initial state results in a set of reduced states, and iii) the decoder (blue arrows), which maps the reduced states to high-dimensional approximate states.}
	\label{fig:framework}
\end{figure}

With all these neural networks defined, the forward pass of the framework can be described as 
\begin{enumerate}
\item encode a reduced initial state from the given initial condition: $\initCondReduced = \encoder(\initCond;\encParams)$,
\item solve a system of ODEs defined by PNODE (or NODE):
$$\reducedSolution^{1},\ldots,\reducedSolution^{\nSteps} = \text{ODESolve}(\initCondReduced, \pnodeVelocity, \PDEparams, \timeSymb_1,\ldots,\timeSymb_{\nSteps}),$$
\item decode a set of reduced states to a set of high-dimensional approximate states: $\approxSolution^{k}= \decoder(\reducedSolution^{k};\decParams), k=1,\ldots,\nSteps$, and
\item compute a loss function $\lossSymb(\approxSolution^{1},\ldots,\approxSolution^{\nSteps},\solution^{1},\ldots,\solution^{\nSteps})$.
\end{enumerate}
Figure \ref{fig:framework} illustrates the computational graph of the forward pass in the proposed framework. We emphasize that the proposed framework only takes the initial states from the training data and the problem-specific ODE parameters $\PDEparams$ as an input. PNODEs still can learn multiple trajectories, which are characterized by the ODE parameters, even if the same initial states are given for different ODE parameters, which is not achievable with NODEs. Furthermore, the proposed framework is significantly simpler than the common neural network settings for NODEs when they are used to learn latent dynamics: the sequence-to-sequence architectures as in \cite{chen2018neural, rubanova2019latent,yildiz2019ode2vae,maulik2020reduced,maulik2020time}, which require that a (part of) sequence is fed into the encoder network to produce a context vector, which is then fed into the NODE decoder network as an initial condition. 

\section{Numerical experiments}
In the following, we apply the proposed framework for learning latent dynamics of  parameterized dynamics from computational physics problems. We then demonstrate the effectiveness of the proposed framework with results of numerical experiments performed on these benchmark problems.

\subsection{Data collection} To train the proposed framework, we collect snapshots of reference solutions by solving the FOM for pre-specified training parameter instances $\PDEparams \in \paramSpaceTrain \equiv \{\PDEparamsTrain^k\}_{k=1}^{\nTrain} \subset \paramSpaceSymb$. This collection results in a tensor $$\solutionTensor \in \RR{\nTrain \times (\nSteps+1) \times \dofFOM},$$ where $\nSteps$ is the number of time steps. The mode-2 unfolding \cite{kolda2009tensor} 
of the solution tensor $\solutionTensor$ gives 
$$\solutionMatrix_{[2]} = \left[ \solutionMatrix(\PDEparamsTrain^1) \;  \cdots \;  \solutionMatrix(\PDEparamsTrain^{\nTrain})\right] \in \RR{\dofFOM \times \nTrain(\nSteps+1)},$$ 
where $\solutionMatrix(\PDEparamsTrain^k) \in \RR{\dofFOM\times(\nSteps+1)}$ consists of the FOM solution snapshots for $\PDEparamsTrain^k$ and the first column corresponds to the initial condition $\solution^{0}(\PDEparamsTrain^k)$. Among the collected solution snapshots, only the first columns of $\solutionMatrix(\PDEparamsTrain^k),k=1,\ldots,\nTrain$ (i.e., the initial conditions) are fed into the framework, the rest of solution snapshots are used in computing the loss function.

Assuming the FOM arises from a spatially discretized partial differential equation, 
the total degrees of freedom $\dofFOM$ can be defined as $\dofFOM = \dofVars \times \dofFOMx \times \cdots \times n_{n_d}$, where $\dofVars$ is the number of different types of solution variables (e.g., chemical species), and $n_{n_d}$ denotes the number of spatial dimensions of the partial differential equation.
 Note that this spatially-distributed data representation is analogous to multi-channel images (i.e., $\dofVars$ corresponds to the number of channels); as such we utilize (transposed) convolutional layers \cite{lecun2015deep,goodfellow2016deep} in our encoder and decoder.

\subsection{Network architectures, training, and testing} In the experiments, we employ convolutional encoders and transposed-convolutional decoders. The encoder consists of four convolutional layers, followed by one fully-connected layer, and the decoder consists of one fully-connected layer, followed by four transposed-convolutional layers. To decrease/increase the spatial dimension, we employ strides larger than one, but we do not use pooling layers. For the nonlinear activation functions, we use ELU \cite{clevert2015fast} after each (transposed) convolutional layer and fully-connected layer, with an exception of the output layer (i.e., no activation at the output layer). 

Moreover, we employ NODE and PNODE for learning latent dynamics and model $\pnodeVelocity$ as fully-connected layers. For the nonlinear activation functions, we again use ELU. Lastly, for $\text{ODESolve}$, we use the Dormand--Prince method \cite{dormand1980family}, which is provided in the software package of \cite{chen2018neural}. Our implementation reuses many parts of the software package used in \cite{rubanova2019latent}, which is written in \texttt{PyTorch}. The details of the configurations will be presented in each of the benchmark problems. 

For training, we set the loss function as the mean squared error and optimize the network weights $(\NODEParams, \encParams, \decParams)$ using Adamax, a variant of Adam \cite{kingma2014adam}, with an initial learning rate 1e-2. At each epoch of training, the loss function is evaluated on the validation set, and the best performing network weights on the validation set are chosen to try the model on the test data. The details on a training/validating/testing-split will be given later in the descriptions of each experiment.

For the performance evaluation metrics, we measure errors of approximated solutions with respect to the reference solutions for testing parameter instances, $\PDEparamsTest^k$. We use the relative $\ell^2$-norm of the error: 
\begin{equation}\label{eq:rel_error}
\frac{\| \solutionMatrix(\PDEparamsTest^k) - \approxSolutionMatrix(\PDEparamsTest^k)  \|\fro}{\| \solutionMatrix(\PDEparamsTest^k) \|\fro},
\end{equation}
where $\|\cdot\|\fro$ denotes the Frobenius norm.

\begin{figure}[!h]
    \centering
    \subfloat[$\timeSymb=7.77$]    {\includegraphics[scale=.5]{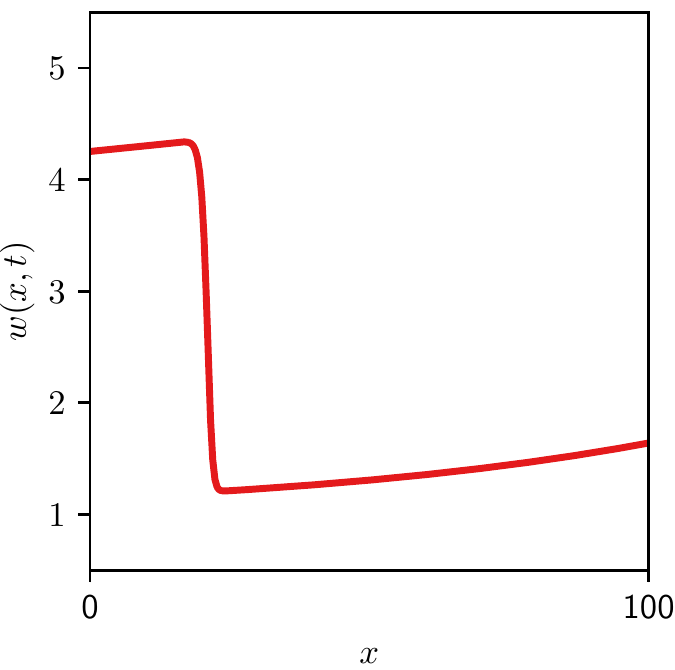}}
    \subfloat[$\timeSymb=11.7$]    {\includegraphics[scale=.5]{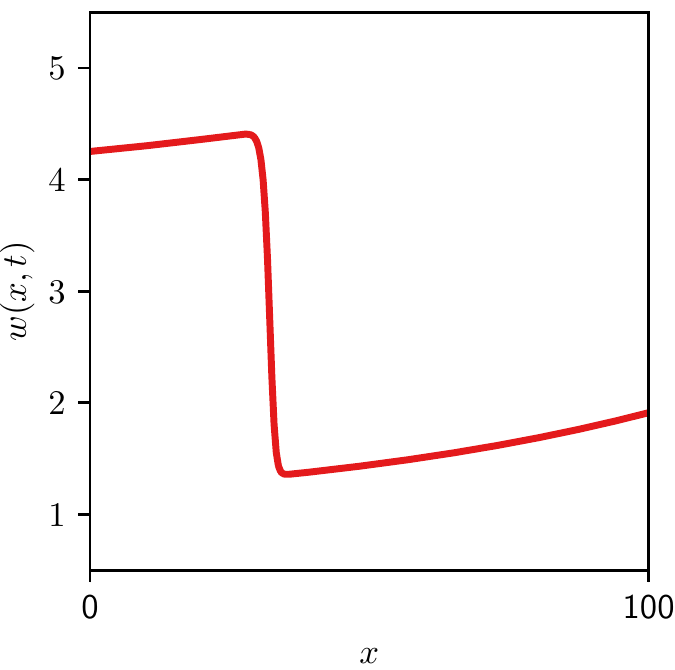}}
    \subfloat[$\timeSymb=19.5$]    {\includegraphics[scale=.5]{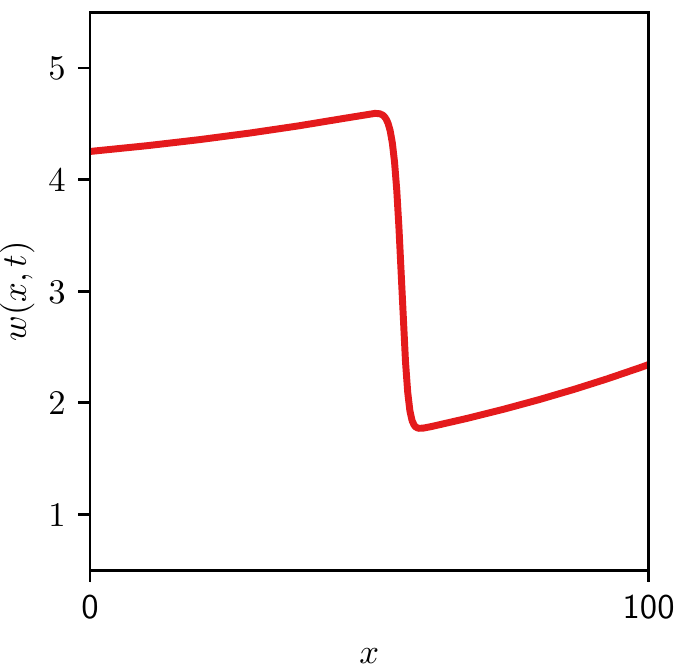}}\\
    \subfloat[$\timeSymb=23.3$]    {\includegraphics[scale=.5]{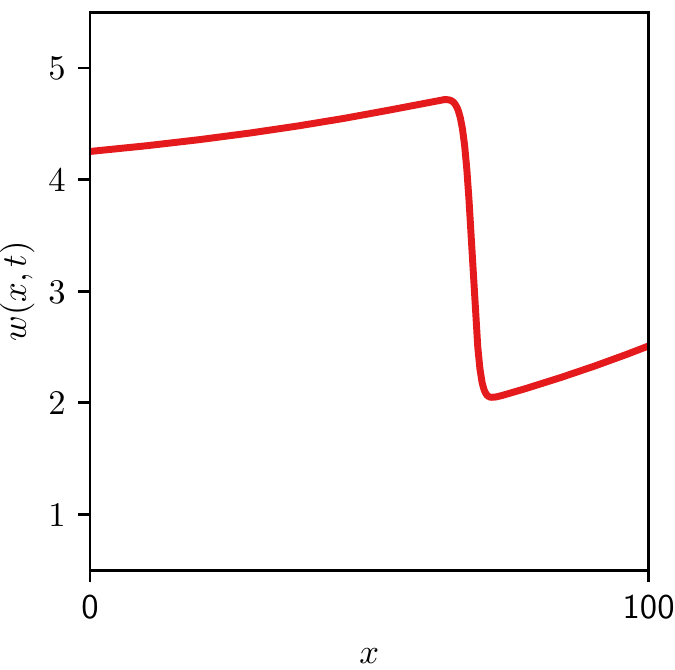}}
    \subfloat[$\timeSymb=27.2$]    {\includegraphics[scale=.5]{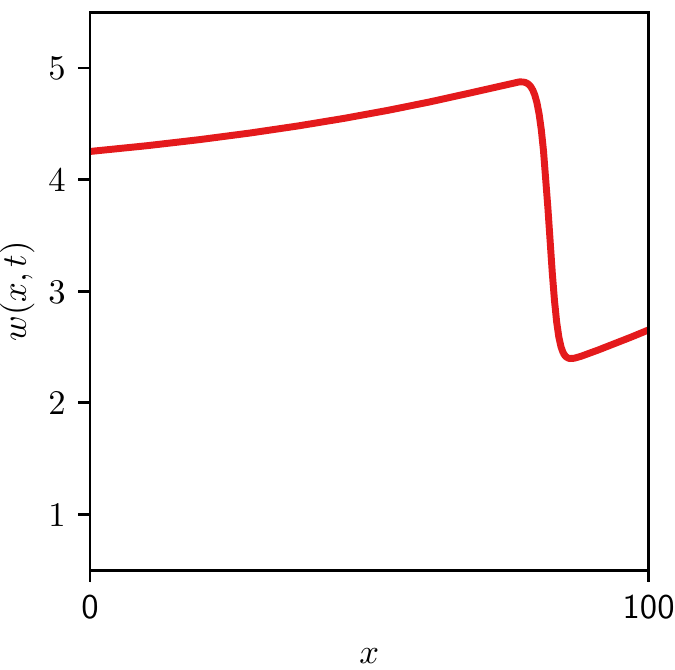}}
    \subfloat[$\timeSymb=35.0$]    {\includegraphics[scale=.5]{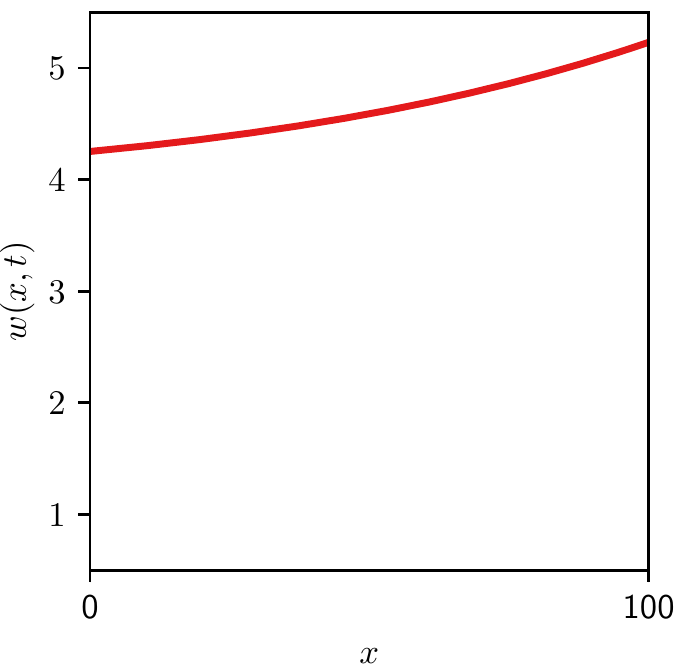}}
    \caption{Snapshots of reference solutions at $\timeSymb=\{7.77,11.7,19.5,23.3,27.2,35.0\}$.}
	\label{fig:burg_ref_sol}
\end{figure}

\subsection{Problem 1: 1D inviscid Burgers' equation}
The first benchmark problem is a parameterized one-dimensional inviscid Burgers' equation, which models simplified nonlinear fluid dynamics and demonstrates propagations of shock. The governing system of partial differential equations is  
\begin{equation}\label{eq:burg_govern}
\frac{\partial \primSolutionSymb(\positionalSymb,\timeSymb;\PDEparams)}{\partial \timeSymb} + \frac{\partial f(\primSolutionSymb(\positionalSymb,\timeSymb;\PDEparams))}{\partial \positionalSymb} = 0.02 e^{\paramSymb_2 \positionalSymb},
\end{equation}
where $f(\primSolutionSymb) = 0.5 \primSolutionSymb^2$, $\positionalSymb \in [0,100]$, and $\timeSymb \in [0,35]$. The boundary condition $\primSolutionSymb(0,\timeSymb;\PDEparams)=\paramSymb_1$ is imposed on the left boundary ($\positionalSymb=0$) and the initial condition is set by $\primSolutionSymb(\positionalSymb,0;\PDEparams)=1$.
Thus, the problem is characterized by a single variable $\primSolutionSymb$ (i.e., $\dofVars=1$) and the two parameters $(\paramSymb_1,\paramSymb_2)$ (i.e., $\nParams=2$) which correspond to the Dirichlet boundary condition at $x=0$ and the forcing term, respectively. Following \cite{Rewienski2003trajectory}, discretizing Eq.~\ref{eq:burg_govern} with Godunov's scheme with 256 control volumes results in a system of parameterized ODEs (FOM) with $\dofFOM=\dofVars \dofFOMx=256$. We then solve the FOM using the backward-Euler scheme 
with a uniform time step $\Delta \timeSymb=0.07$, which results in $\nSteps=500$ for each $\PDEparamsTrain^k \in \paramSpaceTrain$. Figure \ref{fig:burg_ref_sol} depicts snapshots of reference solutions for parameter instances $(\paramSymb_1,\paramSymb_2) = (4.25, 0.015)$ at time $\timeSymb=\{7.77,11.7,19.5,23.3,27.2,35.0\}$, illustrating the discontinuity (shock) moving from left to right as time proceeds.

For the numerical experiments in this subsection, we use the network described in Table \ref{tab:burg_network_architecture}.

\begin{table}[!h]
{\footnotesize
  \caption{Network architecture: kernel filter length $\kernelLength$, number of kernel filters $\nKernels$, and strides $\strides$ at each (transposed) convolutional layers.}\label{tab:burg_network_architecture}
\begin{center}
\renewcommand{\arraystretch}{1.25}
  \begin{tabular}{|c|c|} 
\multicolumn{2}{c}{Encoder}\\
\hline
\multicolumn{2}{|c|}{Conv-layer (4 layers)}\\
\hline
$\kernelLength$  &[16, \phantom{1}8, \phantom{1}4, \phantom{1}4] \\
$\nKernels$ &[\phantom{1}8, 16, 32, 64] \\
$\strides$ &[\phantom{1}2, \phantom{1}4, \phantom{1}4, \phantom{1}4] \\
\hline
\multicolumn{2}{|c|}{FC-layer (1 layer)}\\
\hline
\multicolumn{2}{|c|}{$\inputDim=128$, $\outputDim=\dofROM$}\\
\hline
\end{tabular}
\hspace{5mm}
\begin{tabular}{|c|c|} 
\multicolumn{2}{c}{Decoder}\\
\hline
\multicolumn{2}{|c|}{FC-layer (1 layer)}\\
\hline
\multicolumn{2}{|c|}{$\inputDim=\dofROM$, $\outputDim=128$}\\
\hline
\multicolumn{2}{|c|}{Trans-conv-layer (4 layers)}\\
\hline
$\kernelLength$  &[\phantom{1}4, \phantom{1}4, \phantom{1}8, 16] \\
$\nKernels$ &[32, 16, \phantom{1}8, \phantom{1}1] \\
$\strides$ &[\phantom{1}4, \phantom{1}4, \phantom{1}4, \phantom{1}2] \\
\hline
\end{tabular}
\end{center}
}
\end{table}

\subsubsection{Reconstruction: approximating a single trajectory with latent-dynamics learning}
\begin{figure}[!h]
    \centering
    \subfloat[NODE]    {\includegraphics[scale=.7]{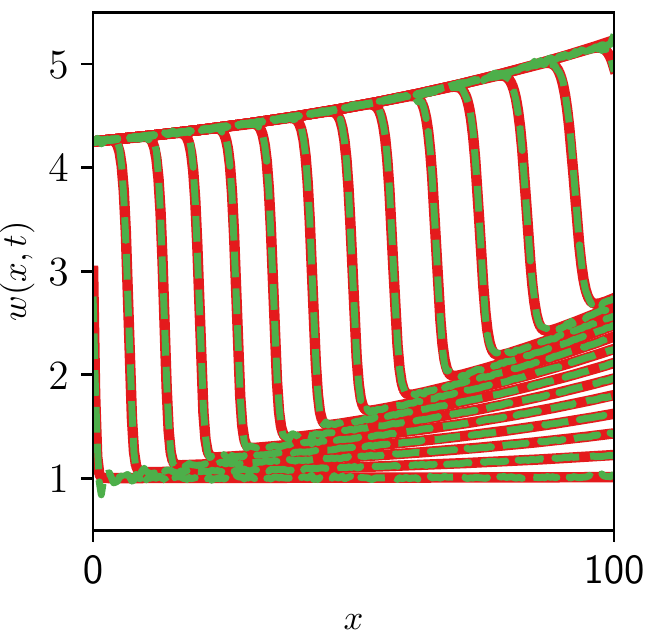} \label{fig:burg_recon_node}} 
    \subfloat[PNODE]    {\includegraphics[scale=.7]{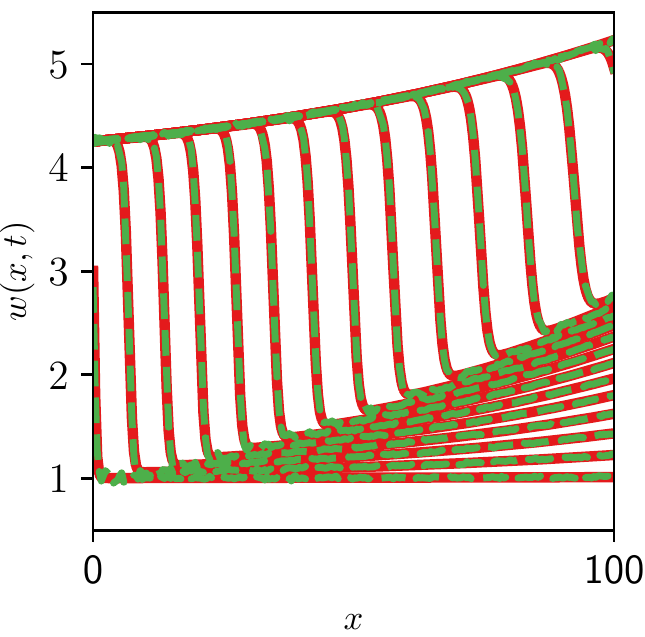}\label{fig:burg_recon_pnode}}
    \caption{Reconstruction: snapshots of reference solutions and approximated solutions using NODE (left) and PNODE (right) at $\timeSymb=\{\frac{35}{15}k\}_{k=1}^{15}$.}
	\label{fig:burg_recon_sol}
\end{figure}

In this experiment, we consider a single training/testing parameter instance $\PDEparamsTrain^1=\PDEparamsTest^1=(\PDEparams_1^1,\PDEparams_2^1)=(4.25,0.015)$ and test both NODE and PNODE for latent-dynamics modeling. We set the reduced dimension\footnote{For setting $\dofROM$, we follow the results of the study on the effective latent dimension shown in \cite{lee2019deep}.} as $\dofROM=5$ and the maximum number of epoch as 20,000. Figure \ref{fig:burg_recon_sol} depicts snapshots of the reference solutions and approximated solutions computed by using the framework with NODE (Figure \ref{fig:burg_recon_node}) and with PNODE (Figure \ref{fig:burg_recon_pnode}) at 15 time instances $\timeSymb=\{\frac{35}{15}k\}_{k=1}^{15}$. 
The relative errors (Eq.~\ref{eq:rel_error}) for NODE and PNODE are $2.6648\times 10^{-3}$ and $ 2.6788\times 10^{-3}$; the differences between the two errors are negligible.

\subsubsection{Prediction: approximating an unseen trajectory for unseen parameter instances}
\begin{figure}[!h]
    \centering
    \subfloat[Scenario 1] {\includegraphics[scale=.8]{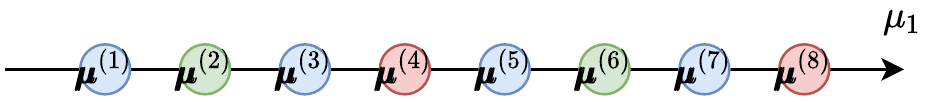} \label{fig:param1_grid}} \\
    \subfloat[Scenario 2] {\includegraphics[scale=.8]{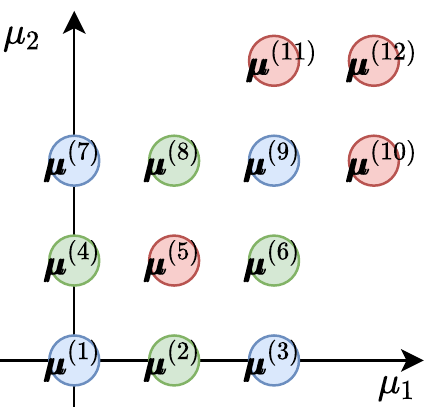} \label{fig:param12_grid}} \hspace{2.5mm}
     {\includegraphics[scale=.8]{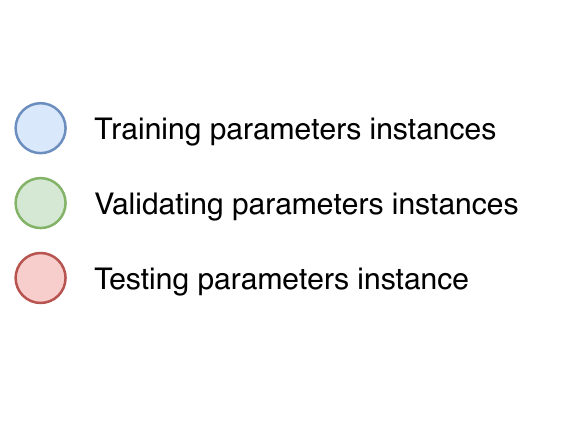} }
    \caption{Visualizations of parameter instances sampling for Scenario 1 and Scenario 2.}
	\label{fig:param_grids}
\end{figure}

We now consider two multi-parameter scenarios. In the first scenario (Scenario 1), we vary the first parameter (boundary condition) and consider 4 training parameter instances, 2 validation parameter instances, and 2 test parameter instances. The parameter instances are collected as shown in Figure \ref{fig:param1_grid}: the training parameter instances correspond to $\paramSpaceTrain=\{(4.25+(0.139)k, 0.015)\}, k=0,2,4,6$, the validating parameter instances correspond to $\paramSpaceVal=\{(4.25+(0.139)k, 0.015)\}, k=1,5$, and the testing parameter instances correspond to $\paramSpaceTest=\{(4.67,0.015), (5.22,0.015)\}$.  
Note that the initial condition is identical for all parameter instances, i.e., $\initCond=1$.

\begin{figure}[!h]
    \centering
    \subfloat[NODE, $\PDEparamsTest^1$]    {\includegraphics[scale=.7]{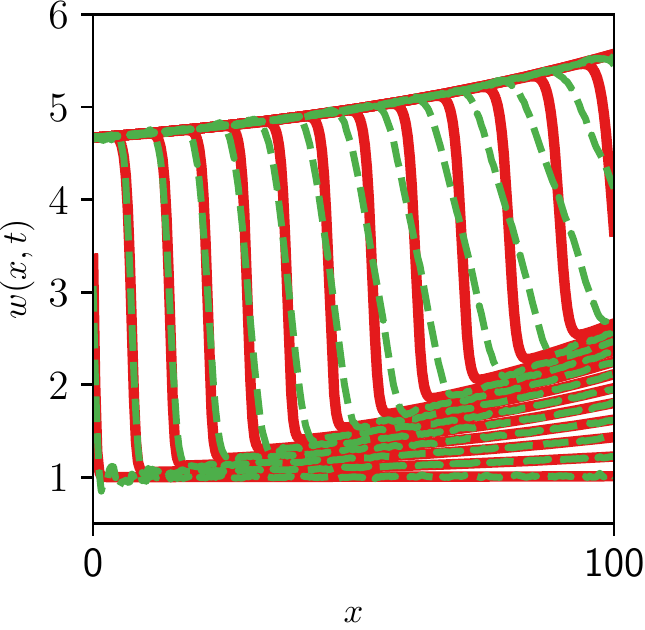} \label{fig:burg_pred_node1}} 
    \subfloat[PNODE, $\PDEparamsTest^1$]    {\includegraphics[scale=.7]{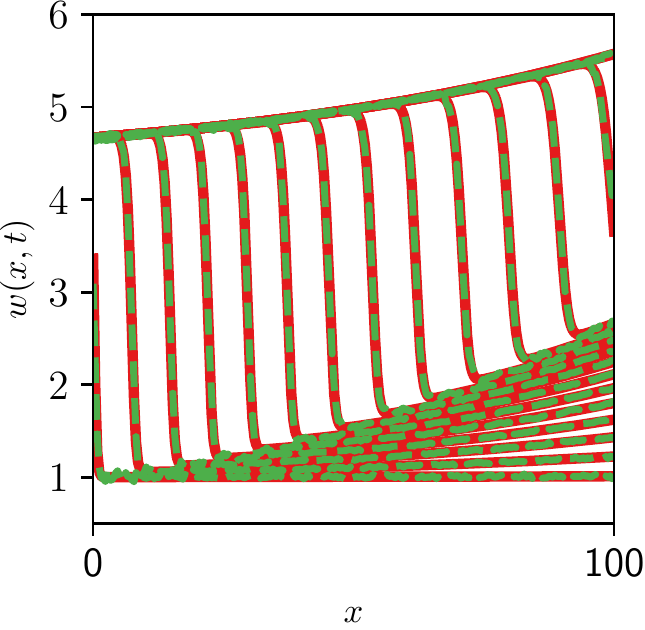}\label{fig:burg_pred_pnode1}}\\
    \subfloat[NODE, $\PDEparamsTest^2$]    {\includegraphics[scale=.7]{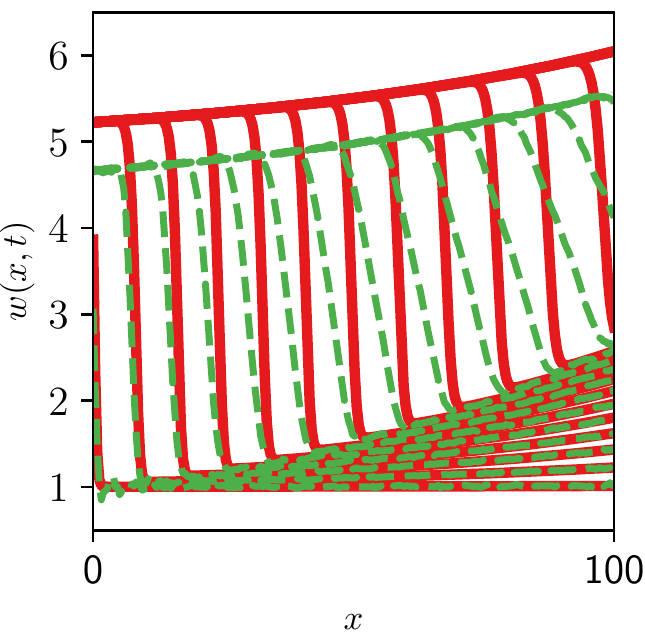} \label{fig:burg_pred_node2}} 
    \subfloat[PNODE, $\PDEparamsTest^2$]    {\includegraphics[scale=.7]{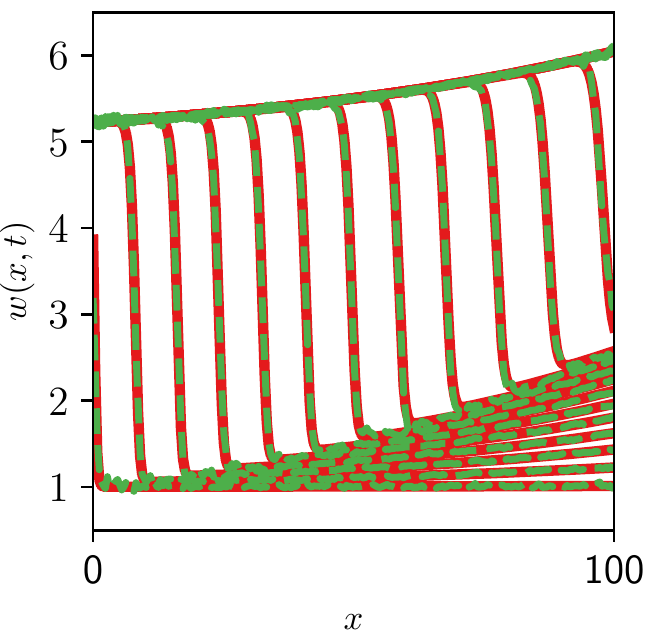}\label{fig:burg_pred_pnode2}}
    \caption{Prediction Scenario 1: snapshots of reference solutions (red) and approximated solutions (green) using NODE (left) and PNODE (right) at $\timeSymb=\{\frac{35}{15}k\}_{k=1}^{15}$ for $\PDEparamsTest^1=(4.67,0.015)$ (top) and $\PDEparamsTest^2=(5.22,0.015)$ (bottom).}
	\label{fig:burg_pred1_sol}
\end{figure}

We train the framework with NODE and PNODE with the same set of hyperparameters with the maximum number of epochs as 50,000. Again, the reduced dimension is set to $\dofROM=5$. Figures \ref{fig:burg_pred_node1}--\ref{fig:burg_pred_pnode1} depict snapshots of reference solutions and approximated solutions using NODE and PNODE.  Both NODE and PNODE learn the boundary condition (i.e., 4.67 at $\positionalSymb=0$) accurately. For NODE, this is only because the testing boundary condition is linearly in the middle of two validating boundary conditions (and also in the middle of four training boundary conditions) and minimizing the mean squared error results in learning a single trajectory with the NODE, where the trajectory has a boundary condition, which is exactly the middle of two validating boundary conditions $4.389$ and $4.944$. Moreover, as NODE learns a single trajectory that minimizes MSE, it actually fails to learn the correct dynamics and results in poor approximate solutions as time proceeds. As opposed to NODE, the PNODE accurately approximates solutions up to the final time. Table \ref{tab:burg_prediction1_error} (second row) shows the relative $\ell^2$-errors (Eq.~\ref{eq:rel_error}) for both NODE and PNODE.

Continuing from the previous experiment, we test the second testing parameter instance,  $\paramSpaceTest=\{(5.22,0.015)\}$, which is located outside $\paramSpaceTrain$ (i.e., next to $\PDEparams^{(7)}$ in Figure \ref{fig:param1_grid}). 
The results are shown in Figures \ref{fig:burg_pred_node2}--\ref{fig:burg_pred_pnode2}: the NODE only learns a single trajectory with the boundary condition, which lies in the middle of validating parameter instances, whereas the PNODE accurately produces approximate solutions for the new testing parameter instances. Table \ref{tab:burg_prediction1_error} (third row) reports the relative errors.

\begin{table}[!h]
{\footnotesize
  \caption{Prediction Scenario 1: the relative $\ell^2$-errors (Eq~\ref{eq:rel_error}).}\label{tab:burg_prediction1_error}
\begin{center}
\renewcommand{\arraystretch}{1.25}
  \begin{tabular}{|c|c|c|} 
  \hline
  & NODE & PNODE\\
  \hline
  $\PDEparamsTest^1=\PDEparams^{(4)}$ & $4.3057\times 10^{-2}$ & $3.6547\times10^{-3}$\\
  \hline
  $\PDEparamsTest^2=\PDEparams^{(8)}$ & $1.5740\times 10^{-1}$ & $5.6900\times10^{-3}$\\
  \hline
\end{tabular}
\end{center}
}
\end{table}

Next, in the second scenario (Scenario 2), we vary both parameters $\paramSymb_1$ and $\paramSymb_2$ as shown in Figure \ref{fig:param12_grid}: the sets of the training, validating, and testing parameter instances correspond to 
\begin{align*}
\paramSpaceTrain&=\{(4.25+(0.139)k, 0.015+(0.002)l)\}, \{(k,l)\}=\{(0,0),(0,2),(2,0),(2,2)\}, \\
\paramSpaceVal&=\{(4.25+(0.139)k, 0.015+(0.002)l)\}, \{(k,l)\}=\{(1,0),(0,1),(2,1),(1,2)\}, \\
\paramSpaceTest&=\{(4.25 + (0.139)k,0.0015 + (0.002)l )\}, \{(k,l)\} = \{(1,1),(3,2),(2,3),(3,3)\}.
\end{align*}

We have tested the set of testing parameter instances and Table \ref{tab:burg_prediction2_error} reports the relative errors; the result shows that PNODE achieves sub 1\% error in most cases. On the other hand, NODE achieves around 10\% errors in most cases. The 1.7\% error of NODE for $\PDEparamsTest^1$ is achieved only because the testing parameter instance is located in the middle of the validating parameter instances (and the training parameter instances).


\begin{table}[!h]
{\footnotesize
  \caption{Prediction Scenario 2: the relative $\ell^2$-errors (Eq~\ref{eq:rel_error}) of the approximate solutions for testing parameter instances.}\label{tab:burg_prediction2_error}
\begin{center}
\renewcommand{\arraystretch}{1.25}
  \begin{tabular}{|l|c|c|} 
  \hline
  & NODE & PNODE\\
  \hline
  $\PDEparamsTest^1=\PDEparams^{(5)}$ & $1.7422\times 10^{-2}$ & $3.2672\times10^{-3}$\\
  \hline
  $\PDEparamsTest^2=\PDEparams^{(10)}$ & $1.0713\times 10^{-1}$ & $7.7303\times10^{-3}$\\
  $\PDEparamsTest^3=\PDEparams^{(11)}$ & $8.9229\times 10^{-2}$ & $8.5650\times10^{-3}$\\  
  $\PDEparamsTest^4=\PDEparams^{(12)}$ & $1.2377\times 10^{-1}$ & $1.0735\times10^{-2}$\\  
  \hline
\end{tabular}
\end{center}
}
\end{table}

\paragraph{Study on effective latent dimension} We further have tested the framework with NODE and PNODE for varying latent dimensions $\dofROM=\{1,2,3,4,5\}$ with the same hyperparameters described in Table~\ref{tab:burg_network_architecture} with the maximum number of epochs as 50,000 and the same training/validating/testing split shown in Figure \ref{fig:param12_grid}. For all testing parameter instances, the dimension of latent states marginally affects the performance of the NODEs. We believe this is because NODE learns a dynamics that minimizes the MSE over four validating parameter instances in $\paramSpaceVal$ regardless of the latent dimensions. On the other hand, decreasing the latent dimension smaller than two ($\dofROM < 3$) negatively affects the performances of the PNODEs for all testing parameter instances. Nevertheless, even with the latent dimension one, $\dofROM=1$, PNODE still outperforms NODE in all testing parameter instances; with $\dofROM=2$, PNODE starts to produce almost order-of-magnitude more accurate approximate solutions than NODE does. Moreover, we observe that, for the given training data/training strategy and the hyperparameters, increasing the latent dimension (larger than $p=2$) only marginally improves the accuracy of the solutions, which agrees with the observations made in \cite{lee2020model} and, in some cases, a neural network work with larger $p$ (i.e., $p>5$) requires more epochs to reach the level of the training accuracy that is achieved by a neural network with smaller $p$ (i.e., $p=\{3,4,5\}$).

\begin{figure}[!h]
    \centering
    \includegraphics[scale=.95]{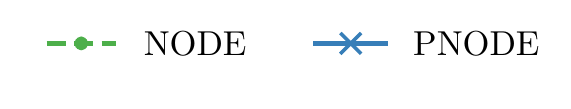}\vspace{-8mm}s\\
    \subfloat[$\PDEparamsTest^1=\PDEparams^{(5)}$]    {\includegraphics[scale=.65]{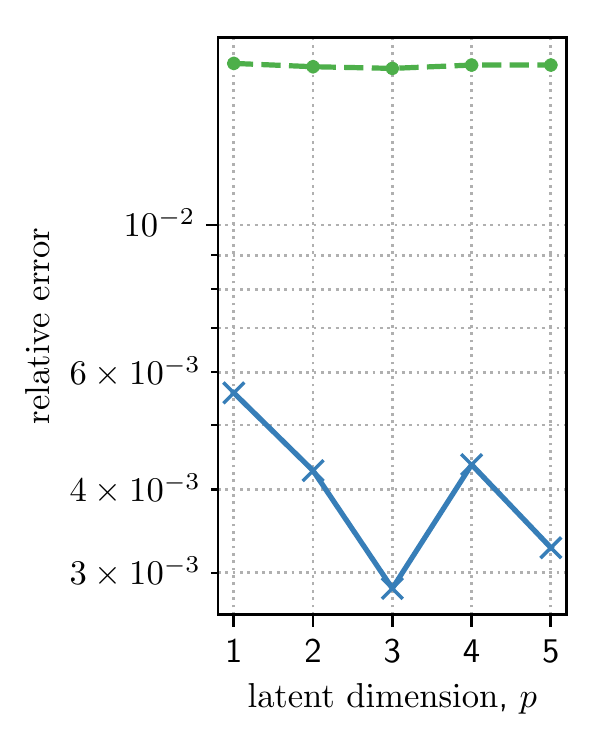}} \vspace{1mm}
    \subfloat[$\PDEparamsTest^2=\PDEparams^{(10)}$]    {\includegraphics[scale=.65]{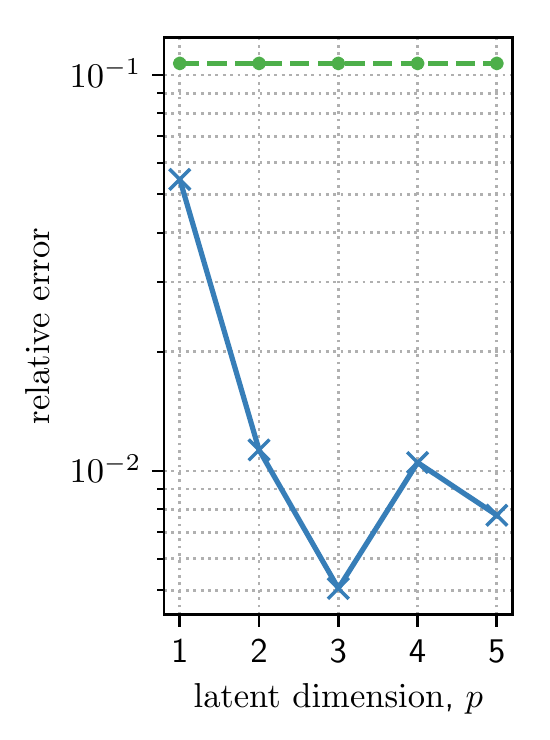}} \vspace{1mm}
    \subfloat[$\PDEparamsTest^3=\PDEparams^{(11)}$]    {\includegraphics[scale=.65]{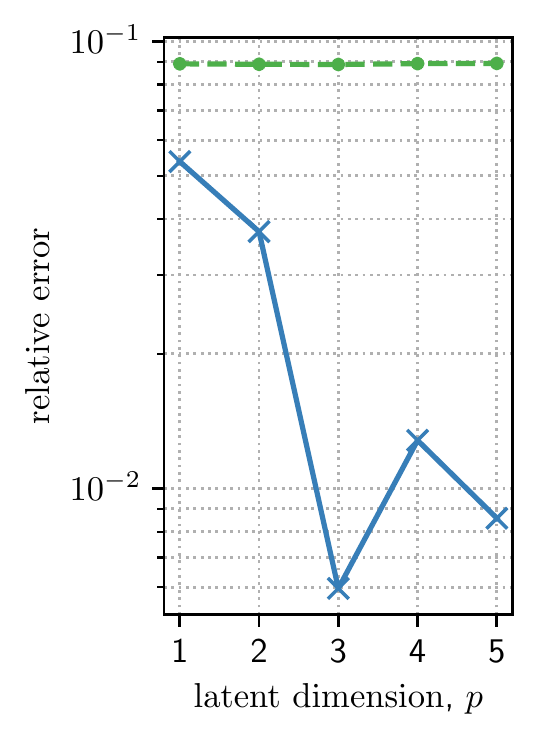}}\vspace{1mm}
    \subfloat[$\PDEparamsTest^4=\PDEparams^{(12)}$]    {\includegraphics[scale=.65]{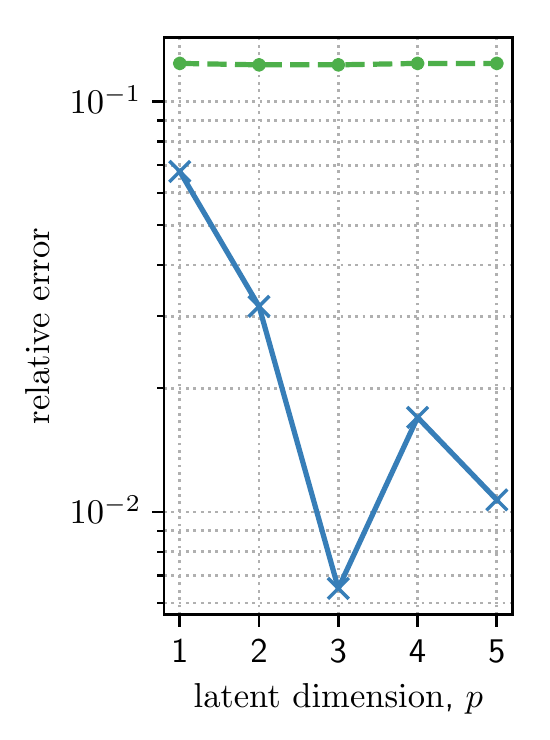}}
    \caption{Prediction Scenario 2: the relative $\ell^2$-errors (Eq~\ref{eq:rel_error}) of the approximate solutions for testing parameter instances with varying latent dimensions $\dofROM=\{1,2,3,4,5\}$.}
	\label{fig:errors}
\end{figure}

\begin{figure}[!h]
    \centering
         {\includegraphics[scale=.8]{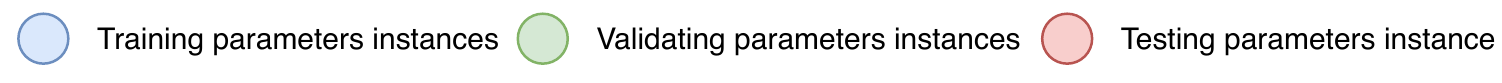}}\\
       \subfloat[Set 1] {\includegraphics[scale=.8]{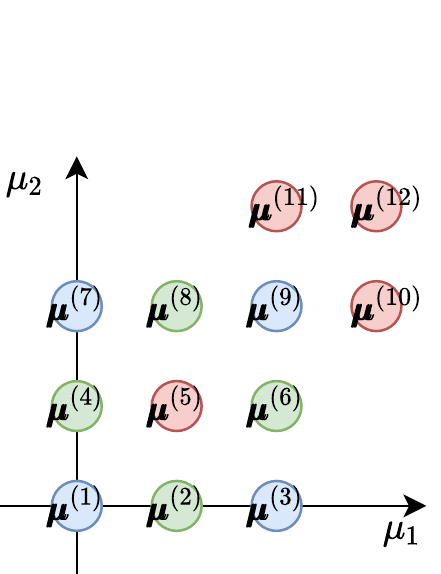} \label{fig:param_var_set1}} \hspace{10mm}
       \subfloat[Set 2] {\includegraphics[scale=.8]{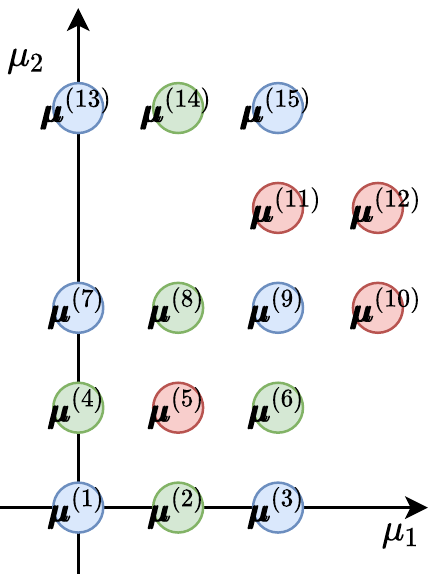} \label{fig:param_var_set2}} \hspace{10mm}
       \subfloat[Set 3] {\includegraphics[scale=.8]{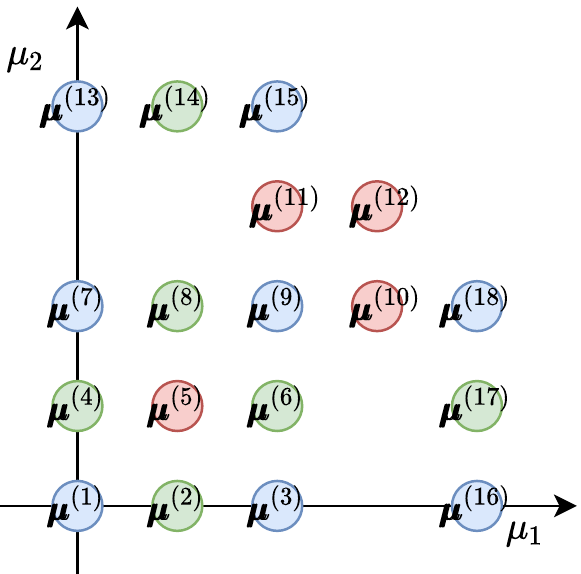} \label{fig:param_var_set3}} 
    \caption{Visualizations of parameter instances sampling for varying number of parameter instances.}
	\label{fig:var_param_grids}
\end{figure}

\paragraph{Study on varying training/validating data sets} We now assess performance of the proposed framework for different settings of training and validating data sets. This experiment illustrates the dependence of the framework on the amount of training data as well as settings of training/validation/testing splits. To this end, we have trained and tested the framework with PNODE on three sets of parameter instance samplings as shown in Figure \ref{fig:var_param_grids}, where the first set in Figure \ref{fig:param_var_set1} is equivalent to the set in Figure \ref{fig:param12_grid}. While all three sets share the testing parameter instances, Sets 2 and 3 (Figures~\ref{fig:param_var_set2} and \ref{fig:param_var_set3}) are built incrementally upon Set 1 by having additional training and validating parameter instances: compared to Set 1, Set 2 has two more training parameter instances $\{\PDEparams^{(13)}, \PDEparams^{(15)}\}$ and one more validating parameter instance $\PDEparams^{(14)}$ as shwon in Figure \ref{fig:param_var_set2} and  Set 3 has four more training parameter instances $\{\PDEparams^{(13)}, \PDEparams^{(15)},\PDEparams^{(16)}, \PDEparams^{(18)}\}$ and two more validating parameter instance $\{\PDEparams^{(14)},\PDEparams^{(17)}\}$ as shwon in Figure \ref{fig:param_var_set3}.

We again consider the same hyperparameters described in Table~\ref{tab:burg_network_architecture} with the maximum number of epochs as 50,000 on the training sets depicted in Figure \ref{fig:var_param_grids}, and Table \ref{tab:burg_var_params} reports the accuracy of approximate solutions computed on the testing parameter instances. Increasing the number of training/validating parameter instances virtually has no effect on the accuracy of the approximation measured on the testing parameter instance $\PDEparams^{(5)}$. That is, for the given network architecture (Table~\ref{tab:burg_network_architecture}), increasing the amount of training/validating data does not have significant effect on the performance of the framework. On the other hand, increasing the number of training/validating parameter instances in a way that is shown in Figure \ref{fig:var_param_grids}  has significantly improved the accuracy of the approximations for the other testing parameter instances $\{\PDEparams^{(10)},\PDEparams^{(11)},\PDEparams^{(12)}\}$. This set of experiments essentially illustrates that, for a given network architecture,  more accurate approximation can be achieved for testing parameter instances that lie in between training/validating parameter instances (i.e., interpolation in the parameter space) than for those that lie outside of training/validating parameter instances (i.e., extrapolation in the parameter space).

\begin{table}[!h]
{\footnotesize
  \caption{Prediction Scenario 2: the relative $\ell^2$-errors (Eq~\ref{eq:rel_error}) of the approximate solutions for testing parameter instances computed using the framework with PNODE. Each framework with PNODE is trained on different training/validating sets depicted in Figure \ref{fig:var_param_grids}.}\label{tab:burg_var_params}
\begin{center}
\renewcommand{\arraystretch}{1.25}
  \begin{tabular}{|l|c|c|c|c|} 
  \hline
  & PNODE -- Set 1 & PNODE -- Set 2 & PNODE -- Set 3  \\
  \hline
  $\PDEparamsTest^1=\PDEparams^{(5)}$ & $3.2672\times10^{-3}$ & $3.3368\times10^{-3}$ & $3.3348\times10^{-3}$\\
  $\PDEparamsTest^2=\PDEparams^{(10)}$ & $7.7303\times10^{-3}$ & $6.6086\times10^{-3}$ & $3.4409\times10^{-3}$\\
  $\PDEparamsTest^3=\PDEparams^{(11)}$ & $8.5650\times10^{-3}$ & $3.7136\times10^{-3}$ & $3.1278\times10^{-3}$\\  
  $\PDEparamsTest^4=\PDEparams^{(12)}$ & $1.0735\times10^{-2}$ & $6.5199\times10^{-3}$& $3.1217\times10^{-3}$\\
  \hline
\end{tabular}
\end{center}
}
\end{table}


\subsection{Problem 2: 2D chemically reacting flow}
The reaction model of a premixed $H_2$-air flame at constant uniform pressure \cite{buffoni2010projection} is described by the equation:
\begin{align}
\frac{\partial \primSolution(\positions,\timeSymb;\PDEparams)}{\partial \timeSymb} &= \nabla \cdot (\diffusionCoeff \nabla\primSolution(\positions,\timeSymb;\PDEparams)) - \bm{v}\cdot\nabla\primSolution(\positions,\timeSymb;\PDEparams) + \sourceTerm(\primSolution(\positions,\timeSymb;\PDEparams);\PDEparams),
\end{align}
where, on the right-hand side, the first term is the diffusion term with the spatial gradient operator $\nabla$, the molecular diffusivity $\diffusionCoeff=2$cm$^2\cdot$s$^{-1}$, the second term is the convective term with the constant and divergence-free velocity field $\bm{v}=[50$cm$\cdot$s$^{-1}$, $0]\tran$, and the third term is the reactive term with the reaction source term $\sourceTerm$. The solution $\primSolution$ corresponds to the thermo-chemical composition vector defined as 
$
\primSolution(\positions,\timeSymb;\PDEparams) = [\primSolutionSymb_{\temperature}(\positions,\timeSymb;\PDEparams),  \primSolutionSymb_{\Htwo}(\positions,\timeSymb;\PDEparams), \primSolutionSymb_{\Otwo}(\positions,\timeSymb;\PDEparams),\primSolutionSymb_{\HtwoO}(\positions,\timeSymb;\PDEparams)]\tran \in \RR{4}
$, where $\primSolutionSymb_{\temperature}$ denotes the temperature, and $\primSolutionSymb_{\Htwo}, \primSolutionSymb_{\Otwo}, \primSolutionSymb_{\HtwoO}$ denote the mass fraction of the chemical species $\Htwo$, $\Otwo$, and $\HtwoO$. The reaction source term is of Arrhenius type, which is defined as: $\sourceTerm(\primSolution;\PDEparams) = [\sourceTermSymb_{\temperature}(\primSolution;\PDEparams), \sourceTermSymb_{\Htwo}(\primSolution;\PDEparams),\sourceTermSymb_{\Otwo}(\primSolution;\PDEparams),\sourceTermSymb_{\HtwoO}(\primSolution;\PDEparams) ]\tran$, where
 \begin{align*}
\sourceTermSymb_{\temperature}(\primSolution;\PDEparams) &= Q \sourceTermSymb_{\HtwoO}(\primSolution;\PDEparams),\\
\sourceTermSymb_{i}(\primSolution;\PDEparams) &= -v_i \left(\frac{W_i}{\rho} \right)\left(\frac{\rho \primSolutionSymb_{\Htwo}}{W_{\Htwo}} \right)^{v_{\Htwo}} \left( \frac{\rho  \primSolutionSymb_{\Otwo}}{W_{\Otwo}}\right)^{v_{\Otwo}} A e^{\left(-\frac{E}{R\primSolutionSymb_{\temperature}}\right)},
\end{align*}
for $i\in\{\Htwo,\Otwo,\HtwoO\}$. Here, $(v_{\Htwo},v_{\Otwo},v_{\HtwoO})=(2,1,-2)$ denote stoichiometric coefficients, $(W_{\Htwo},W_{\Otwo},W_{\HtwoO})=(2.016,31.9,18)$ denote molecular weights in units g$\cdot$mol$^{-1}$, $\rho=1.39\times10^{-3}$g$\cdot$cm$^{-3}$ denotes the density mixture, $R=8.314$J$\cdot$mol$^{-1}\cdot$K$^{-1}$ denotes the universal gas constant, and $Q=9800$K denotes the heat of the reaction. The problem has two input parameters (i.e., $\nParams=2$), which correspond to $\PDEparams=(\paramSymb_1,\paramSymb_2)=(A,E)$, where $A$ and $E$ denote the pre-exponential factor and the activation energy.

\begin{figure}[!h]
    \centering
    {\includegraphics[scale=1]{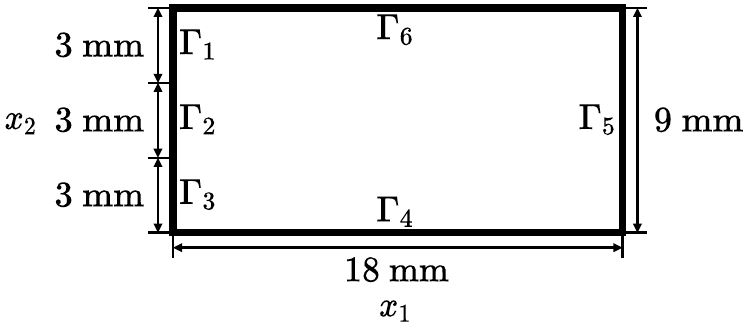}} 
    \caption{The geometry of the spatial domain for chemically reacting flow.}
	\label{fig:adr_domain}
\end{figure}

\begin{figure}[!t]
    \includegraphics[scale=.9]{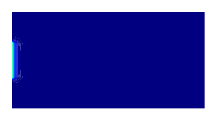}
    \includegraphics[scale=.9]{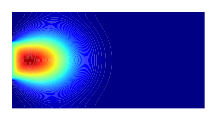}
    \includegraphics[scale=.9]{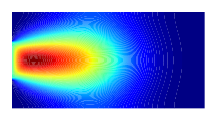}
    \includegraphics[scale=.9]{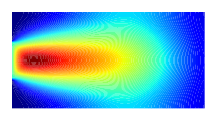}
    \includegraphics[scale=.9]{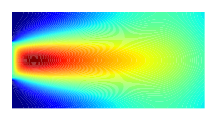}
    \includegraphics[scale=.9]{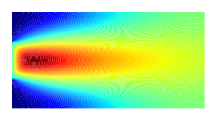}
    \includegraphics[scale=.9]{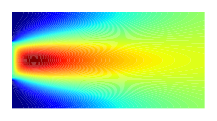}
    \includegraphics[scale=.9]{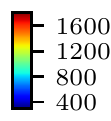}\\
    \includegraphics[scale=.9]{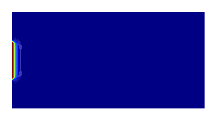}
    \includegraphics[scale=.9]{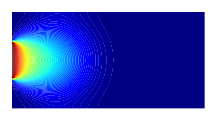}
    \includegraphics[scale=.9]{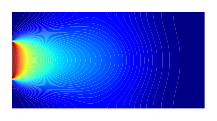}
    \includegraphics[scale=.9]{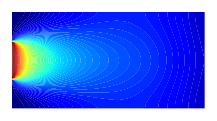}
    \includegraphics[scale=.9]{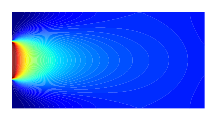}
    \includegraphics[scale=.9]{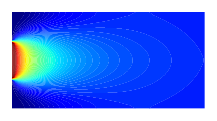}
    \includegraphics[scale=.9]{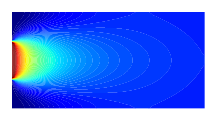}
    \includegraphics[scale=.9]{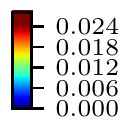}\\
    \includegraphics[scale=.9]{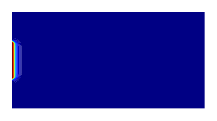}
    \includegraphics[scale=.9]{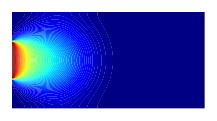}
    \includegraphics[scale=.9]{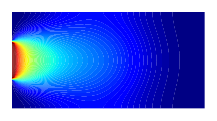}
    \includegraphics[scale=.9]{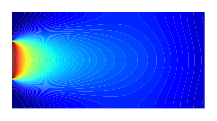}
    \includegraphics[scale=.9]{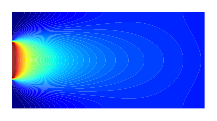}
    \includegraphics[scale=.9]{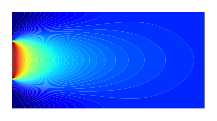}
    \includegraphics[scale=.9]{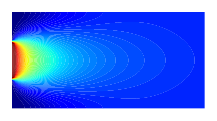}
    \includegraphics[scale=.9]{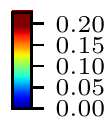}\\
    \includegraphics[scale=.9]{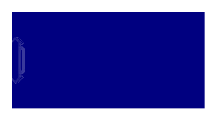}
    \includegraphics[scale=.9]{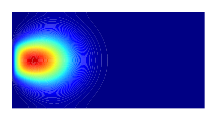}
    \includegraphics[scale=.9]{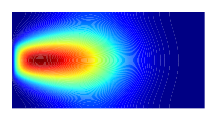}
    \includegraphics[scale=.9]{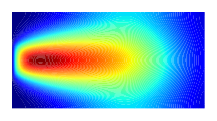}
    \includegraphics[scale=.9]{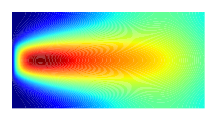}
    \includegraphics[scale=.9]{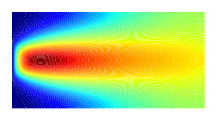}
    \includegraphics[scale=.9]{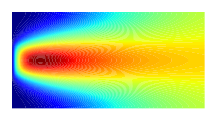}
    \includegraphics[scale=.9]{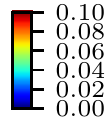}
    \caption{Snapshots of reference solutions of temperature (first row), $\Htwo$ (second row), $\Otwo$ (third row), and $\HtwoO$ (fourth row) at $\timeSymb=\{0,0.01,0.02,0.03,0.04,0.05,0.06\}$ (from left to right).}
	\label{fig:adr_ref_sol}
\end{figure}

Figure \ref{fig:adr_domain} depicts the geometry of the spatial domain and the boundary conditions are set as:
\begin{itemize}
\item $\Gamma_2$: the inflow boundary with Dirichlet boundary conditions $\primSolutionSymb_{\temperature}=950$K, and $(\primSolutionSymb_{\Htwo},\primSolutionSymb_{\Otwo},\primSolutionSymb_{\HtwoO})=(0.0282,0.2259,0)$,
\item $\Gamma_1$ and $\Gamma_3$: the Dirichlet boundary conditions $\primSolutionSymb_{\temperature}=300$K, and $(\primSolutionSymb_{\Htwo},\primSolutionSymb_{\Otwo},\primSolutionSymb_{\HtwoO})=(0,0,0)$,
\item $\Gamma_4,\Gamma_5,$ and $\Gamma_6$: the homogeneous Neumann condition,
\end{itemize}
and the initial condition is set as $\primSolutionSymb_{\temperature}=300$K, and $(\primSolutionSymb_{\Htwo},\primSolutionSymb_{\Otwo},\primSolutionSymb_{\HtwoO})=(0,0,0)$ (i.e., empty of chemical species). For collecting data, we employ a finite-difference method with $64\times 32$ uniform grid points (i.e., $\dofFOM = \nParams\times\dofFOMx\times\dofFOMy=4\times64\times32$), the second-order backward Euler method (BDF2) 
with a uniform time step $\Delta \timeSymb=10^{-4}$ and the final time $0.06$ (i.e., $\nSteps=600$). Figure \ref{fig:adr_ref_sol} depicts  snapshots of the reference solutions of each species for the training parameter instance $(\paramSymb_1,\paramSymb_2) = (2.3375\times10^{12},5.6255\times10^{3})$.

\subsubsection{Data preprocessing and training}
Each species in this problem has different numeric scales: the magnitude of $\primSolutionSymb_{\temperature}$ is about four-orders of magnitude larger than those of other species (see Figure \ref{fig:adr_ref_sol}).  To set the values of each species in a common range $[0, 1]$, we employ zero-one scaling to each species separately. Moreover, because the values of $A$ and $E$ are several orders of magnitude larger than those of species, we scale the input parameters as well to match the scales of the chemical species. We simply divide the values of the first parameter and the second parameter by $10^{13}$ and $10^{4}$, respectively.\footnote{We have not investigated different scaling strategies for scaling of parameter instances as it is not the focus of this study.} After these scaling operations, we train the framework with hyper-parameters specified in Table \ref{tab:adr_network_architecture}, where the input data consists of 2-dimensional data with 4 channels, and the reduced dimension is again set as $\dofROM=5$.

\begin{table}[!h]
{\footnotesize
  \caption{Network architecture: kernel filter length $\kernelLength=\kernelLength_1=\kernelLength_2$, number of kernel filters $\nKernels$, and strides $\strides=\strides_1=\strides_2$ at each (transposed) convolutional layer.}\label{tab:adr_network_architecture}
\begin{center}
\renewcommand{\arraystretch}{1.25}
  \begin{tabular}{|c|c|} 
\multicolumn{2}{c}{Encoder}\\
\hline
\multicolumn{2}{|c|}{Conv-layer (4 layers)}\\
\hline
$\kernelLength$  &[16, \phantom{1}8, \phantom{1}4, \phantom{1}4] \\
$\nKernels$ &[\phantom{1}8, 16, 32, 64] \\
$\strides$ &[\phantom{1}2, \phantom{1}2, \phantom{1}2, \phantom{1}2] \\
\hline
\multicolumn{2}{|c|}{FC-layer (1 layer)}\\
\hline
\multicolumn{2}{|c|}{$\inputDim=512$, $\outputDim=\dofROM$}\\
\hline
\end{tabular}
\hspace{5mm}
\begin{tabular}{|c|c|} 
\multicolumn{2}{c}{Decoder}\\
\hline
\multicolumn{2}{|c|}{FC-layer (1 layer)}\\
\hline
\multicolumn{2}{|c|}{$\inputDim=\dofROM$, $\outputDim=512$}\\
\hline
\multicolumn{2}{|c|}{Trans-conv-layer (4 layers)}\\
\hline
$\kernelLength$  &[\phantom{1}4, \phantom{1}4, \phantom{1}8, 16] \\
$\nKernels$ &[32, 16, \phantom{1}8, \phantom{1}4] \\
$\strides$ &[\phantom{1}2, \phantom{1}2, \phantom{1}2, \phantom{1}2] \\
\hline
\end{tabular}
\end{center}
}
\end{table}

\subsubsection{Prediction: approximating an unseen trajectory for unseen parameter instances}
\begin{figure}[!h]
    \centering
    {\includegraphics[scale=.8]{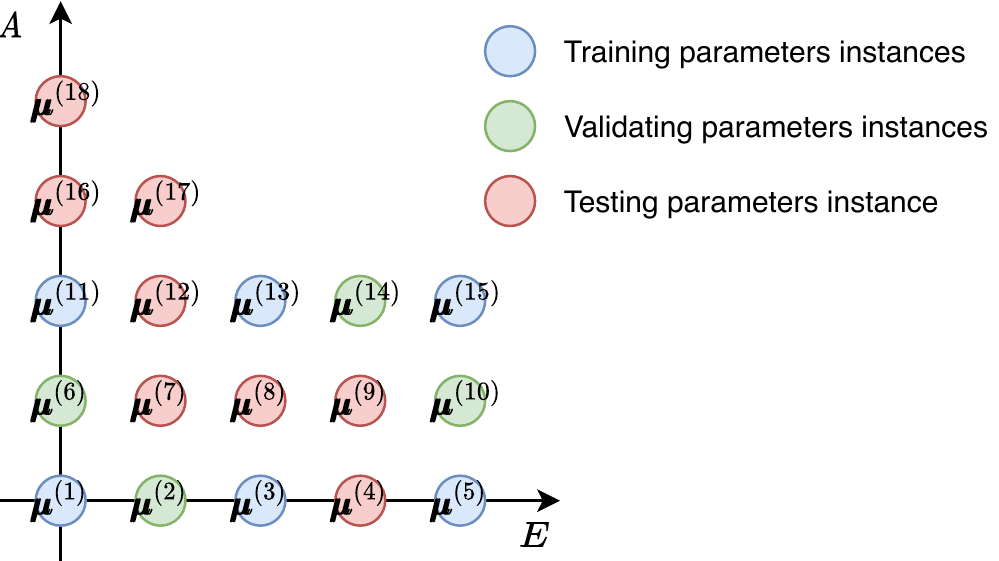}} 
    \caption{Visualizations of parameter instances sampling for the reacting flow.}
	\label{fig:adr_param_grids}
\end{figure}

In this experiment, we vary two parameters: the pre-exponential factor $\paramSymb_1=A$ and the activation energy $\paramSymb_2=E$. We consider parameter instances as depicted in Figure \ref{fig:adr_param_grids}: the sets of the training, validating, and testing parameter instances correspond to 
\begin{alignat*}{2}
\paramSpaceTrain&=\{(2.3375\times 10^{12}+(0.5946\times 10^{12})k, &&5.6255\times 10^{3}+(0.482 \times 10^{3})l)\},  \\
& &&\{(k,l)\}=\{(0,0),(0,2),(0,4),(2,0),(2,2),(2,4)\}, \\
\paramSpaceVal&=\{(2.3375\times 10^{12}+(0.5946\times 10^{12})k, &&5.6255\times 10^{3}+(0.482 \times 10^{3})l)\}, \\
& &&\{(k,l)\}=\{(0,1),(1,0),(1,4),(2,3)\},\\
\paramSpaceTest&=\{(2.3375\times 10^{12}+(0.5946\times 10^{12})k, &&5.6255\times 10^{3}+(0.482 \times 10^{3})l)\},  \\
& &&\{(k,l)\}=\{(0,3),(1,1),(1,2),(1,3),(2,1),(3,0),(3,1),(4,0)\}.
\end{alignat*}


Table \ref{tab:adr_prediction_error} presents the relative $\ell^2$-errors of approximate solutions computed using NODE and PNODE for testing parameter instances in the predictive scenario. The first three rows in Table \ref{tab:adr_prediction_error} correspond to the results of testing parameter instances at the middle three red circles in Figure \ref{fig:adr_param_grids}. As expected, both NODE and PNODE work well for these testing parameter instances: NODE is expected to work well for these testing parameter instances because the single trajectory that minimizes the MSE over validating parameter instances would be the trajectory associated with the testing parameter $\PDEparams^{(8)}$. As we consider testing parameter instances that are distant from $\PDEparams^{(8)}$,  
we observe PNODE to be (significantly) more accurate than NODE.
From these observations, the NODE model can be considered as being overfitted to a trajectory that minimizes the MSE. This overfitting can be avoided to a certain extent by applying e.g., early-stopping, however, this cannot fundamentally fix the problem of the NODE (i.e., fitting a single trajectory for all input data distributions).

\begin{table}[!h]
{\footnotesize
  \caption{Prediction Scenario: the relative $\ell^2$-errors (Eq~\ref{eq:rel_error}).}\label{tab:adr_prediction_error}
\begin{center}
\renewcommand{\arraystretch}{1.25}
  \begin{tabular}{|l|c|c|} 
  \hline
  & NODE & PNODE\\
  \hline
  $\PDEparamsTest^1=\PDEparams^{(7)}$ & $9.2823 \times 10^{-3}$ & $4.2993\times 10^{-3}$\\
  $\PDEparamsTest^2=\PDEparams^{(8)}$ & $3.3450 \times 10^{-3}$ & $4.6429\times 10^{-3}$\\
  $\PDEparamsTest^3=\PDEparams^{(9)}$ & $4.1516 \times 10^{-3}$ & $5.0617\times 10^{-3}$\\  
  \hline
  $\PDEparamsTest^4=\PDEparams^{(4)}$ & $4.0835 \times 10^{-2}$ & $5.6011\times 10^{-3}$\\  
  $\PDEparamsTest^5=\PDEparams^{(12)}$ & $3.4767 \times 10^{-2}$ & $4.4133\times 10^{-3}$\\  
  $\PDEparamsTest^6=\PDEparams^{(16)}$ & $5.9410 \times 10^{-2}$ & $1.2935\times 10^{-2}$\\  
  $\PDEparamsTest^7=\PDEparams^{(17)}$ & $5.4553 \times 10^{-2}$ & $1.1785\times 10^{-2}$\\  
  $\PDEparamsTest^8=\PDEparams^{(18)}$ & $7.4881 \times 10^{-2}$ & $2.4660\times 10^{-2}$\\  
  \hline
\end{tabular}
\end{center}
}
\end{table}

\subsection{Problem 3: Quasi-1D Euler equation}
\begin{figure}[!h]
    \centering
    {\includegraphics[scale=.9]{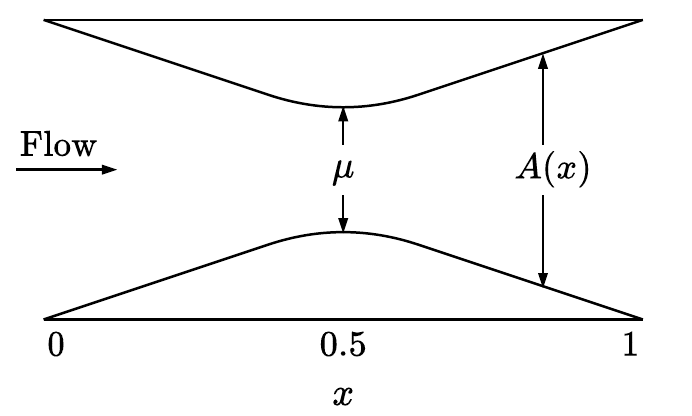} } 
    \caption{The geometry of the spatial domain for quasi-1D Euler equation of converging--diverging nozzle.}
	\label{fig:q1Edomain}
\end{figure}

For the third benchmark problem, we consider the quasi-one-dimensional Euler equations for modeling inviscid compressible flow in a one-dimensional converging--diverging nozzle with a continuously varying cross-sectional area \cite{maccormack2014numerical}. The system of the governing equations is  
\begin{equation*}
\frac{\partial \primSolution}{\partial \timeSymb} + \frac{1}{A}\frac{\partial \forcing(\primSolution)}{\partial \positionalSymb} = \source(\primSolution),
\end{equation*}
where 
\begin{equation*}
\primSolution = \begin{bmatrix}\densitySymb\\ \densitySymb\VelocitySymb\\ \energySymb \end{bmatrix}, \quad \forcing(\primSolution) = \begin{bmatrix} \densitySymb \VelocitySymb\\ \densitySymb \VelocitySymb^2 +  \pressureSymb \\ (\energySymb+\pressureSymb)\VelocitySymb \end{bmatrix}, \quad \source(\primSolution) = \begin{bmatrix} 0 \\ \frac{\pressureSymb}{\areaSymb}\frac{\partial \areaSymb}{\partial \positionalSymb} \\ 0 \end{bmatrix},
\end{equation*}
with $\pressureSymb = (\heatRatio-1)\densitySymb\potentialEnergy$, $\potentialEnergy =\frac{\energySymb}{\densitySymb} - \frac{\VelocitySymb^2}{2}$, and $\areaSymb = \areaSymb(\positionalSymb)$. Here, $\densitySymb$ denotes density, $\VelocitySymb$ denotes velocity, $\pressureSymb$ denotes pressure, $\potentialEnergy$ denotes energy per unit mass, $\energySymb$ denotes total energy density, $\heatRatio$ denotes the specific heat ratio, and $\areaSymb(\positionalSymb)$ denotes the converging--diverging nozzle cross-sectional area. We consider a specific heat ratio of $\heatRatio=1.3$, a specific heat constant of $\gasConstant=355.4 $m$^2$/$s^2$/K, a total temperature of $\totalTemp=300 $K, and a total pressure of $\totalPressure=10^6 $N/m$^2$. The cross-sectional area $\areaSymb(\positionalSymb)$ is determined by a cubic spline interpolation over the points $(\positionalSymb, \areaSymb(\positionalSymb)) = \{(0,0.2), (0.25,1.05\paramSymb), (0.5,\paramSymb), (0.75,1.05\paramSymb),(1,0.2)\}$, where $\paramSymb$ determines the width of the middle cross-sectional area. Figure \ref{fig:q1Edomain} depicts the schematic figures of converging--diverging nozzle determined by $\areaSymb(\positionalSymb)$, parameterized by the width of the middle cross-sectional area, $\paramSymb$. A perfect gas, which obeys the ideal gas law (i.e., $\pressureSymb=\densitySymb \gasConstant \temperatureSymb$), is assumed. 

For the initial condition, the initial flow field is computed as follows; a zero pressure-gradient flow field is constructed via the isentropic relations,
\begin{equation*}
\MachSymb(\positionalSymb) = \frac{\MachSymb\midArea\areaSymb\midArea}{\areaSymb(\positionalSymb)} \left( \frac{1+\frac{\heatRatio-1}{2}\MachSymb(\positionalSymb)^2}{1+\frac{\heatRatio-1}{2}\MachSymb\midArea^2}\right)^{\frac{\heatRatio+1}{2(\heatRatio-1)}}, 
\quad 
\pressureSymb(\positionalSymb) = \totalPressure \left( 1 + \frac{\heatRatio-1}{2}\MachSymb(\positionalSymb)^2\right)^{\frac{-\heatRatio}{\heatRatio-1}}, 
\end{equation*}
\begin{equation*}
\temperatureSymb(\positionalSymb) = \totalTemp \left( 1 + \frac{\heatRatio-1}{2}\MachSymb(\positionalSymb)^2 \right)^{-1}, 
\quad
\densitySymb(\positionalSymb) = \frac{\pressureSymb(\positionalSymb)}{\gasConstant \temperatureSymb(\positionalSymb)},
\quad 
\soundSymb(\positionalSymb) = \sqrt{\gasConstant \frac{\pressureSymb(\positionalSymb)}{\densitySymb(\positionalSymb)}},
\quad
\VelocitySymb(\positionalSymb) = \MachSymb(\positionalSymb) \soundSymb(\positionalSymb),
\end{equation*}
where $\MachSymb$ denotes the Mach number, $\soundSymb$ denotes the speed of sound, a subscript $m$ indicates the flow quantity at $\positionalSymb=0.5$ m. The shock is located at $\positionalSymb=0.85$ m and the velocity across the shock ($\VelocitySymb_2$) is computed by using the jump relations for a stationary shock and the perfect gas equation of state. The velocity across the shock satisfies the quadratic equation 
\begin{equation*}
\left(\frac{1}{2} - \frac{\heatRatio}{\heatRatio-1} \right) \VelocitySymb_2^2 + \frac{\heatRatio}{\heatRatio-1} \frac{n}{m} \VelocitySymb_2 - h = 0, 
\end{equation*}
where $m=\densitySymb_2 \VelocitySymb_2 = \densitySymb_1\VelocitySymb_1, n=\densitySymb_2 \VelocitySymb_2^2 + \pressureSymb_2 = \densitySymb_1\VelocitySymb_1^2 + \pressureSymb_1, h=(\energySymb_2+\pressureSymb_2)/\densitySymb_2 = (\energySymb_1+\pressureSymb_1)/\densitySymb_1$. The subscripts 1 and 2 indicates quantities to the left and to the right of the shock. We consider a specific Mach number of $\MachSymb\midArea=2.0$.

\begin{figure}[!h]
    \centering
    \subfloat[$\timeSymb=0.1$]    {\includegraphics[scale=.5]{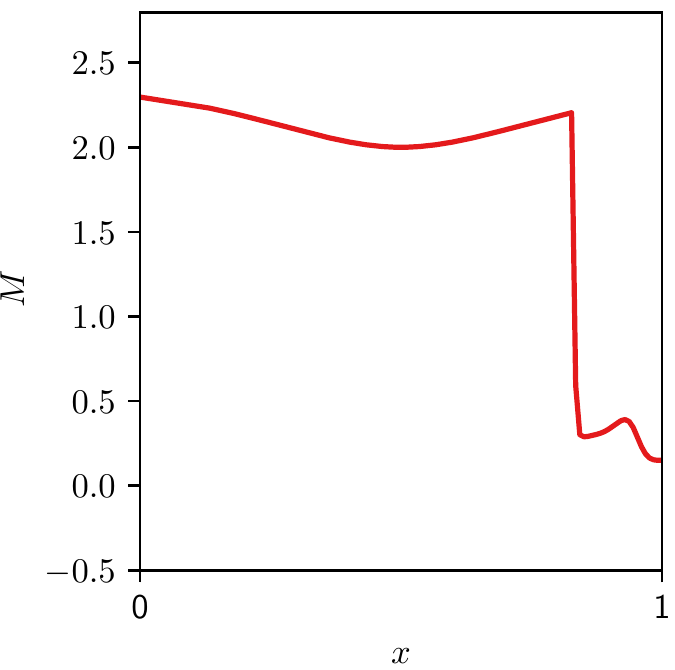}}
    \subfloat[$\timeSymb=0.2$]    {\includegraphics[scale=.5]{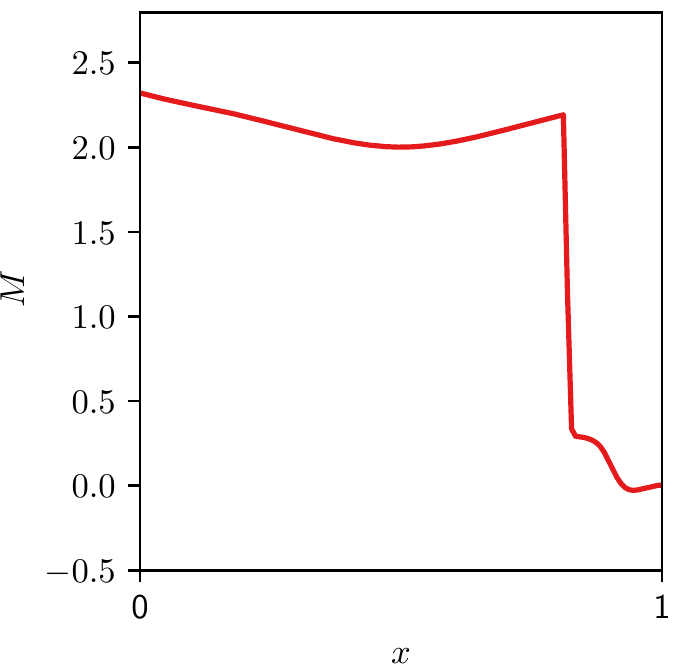}}
    \subfloat[$\timeSymb=0.3$]    {\includegraphics[scale=.5]{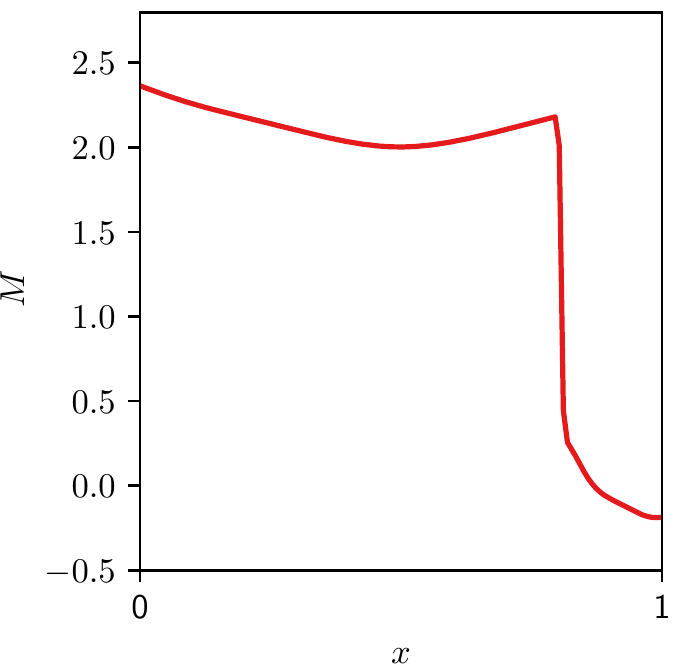}}\\
    \subfloat[$\timeSymb=0.4$]    {\includegraphics[scale=.5]{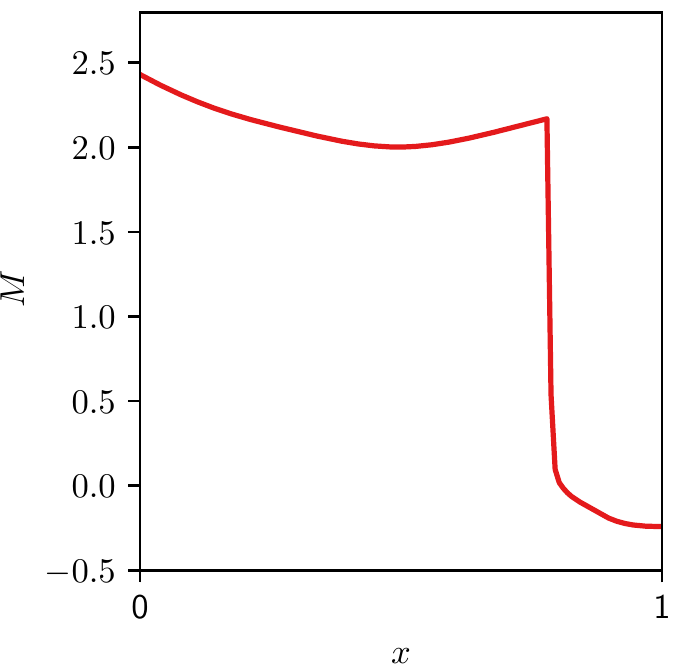}}
    \subfloat[$\timeSymb=0.5$]    {\includegraphics[scale=.5]{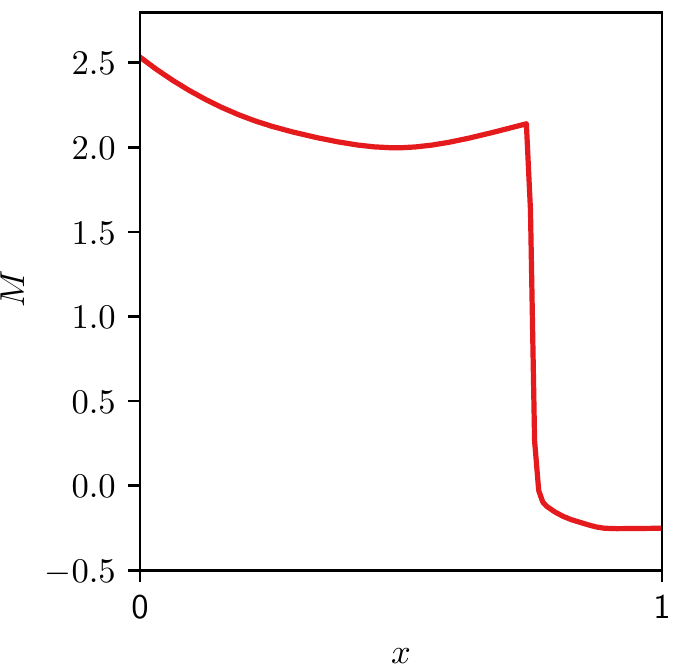}}
    \subfloat[$\timeSymb=0.6$]    {\includegraphics[scale=.5]{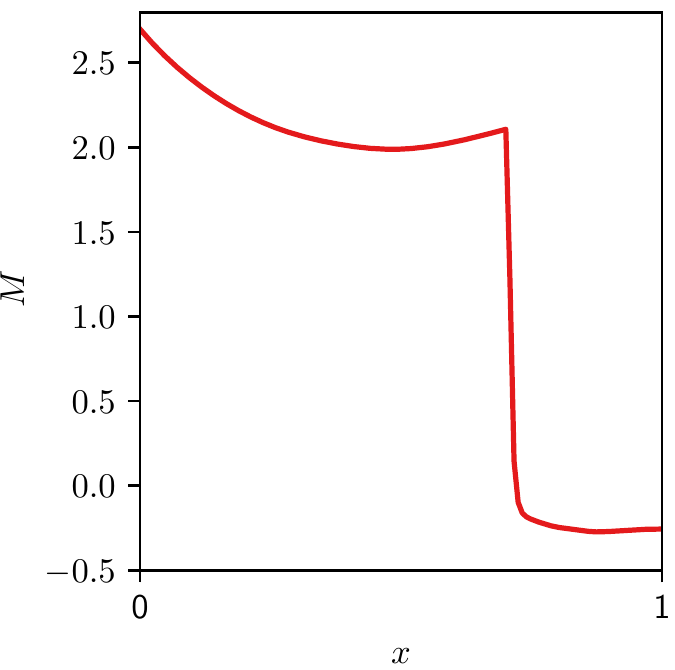}}\\
    \caption{Snapshots of reference solutions of Mach number $\MachSymb(\positionalSymb)$ for $\paramSymb=0.15$ at $\timeSymb=\{0.1,0.2,0.3,0.4,0.5,0.6\}$.}
	\label{fig:euler_ref_sol}
\end{figure}

For spatial discretization, we employ a finite-volume scheme with 128 equally spaced control volumes and fully implicit boundary conditions, which leads to $\dofFOM=\dofVars \dofFOMx =3\times128= 384$. At each intercell face, the Roe flux difference vector splitting method is used to compute the flux. For time discretization, we employ the backward Euler scheme with a uniform time step $\Delta \timeSymb=10^{-3}$ and the final time $0.6$ (i.e., $\nSteps=600$). Figure \ref{fig:euler_ref_sol} depicts the snapshots of reference solutions of Mach number $\MachSymb(\positionalSymb)$ for the middle cross-sectional area $\paramSymb=0.15$ at $\timeSymb=\{0.1,0.2,0.3,0.4,0.5,0.6\}$. 

\begin{figure}[!h]
    \centering
    {\includegraphics[scale=.8]{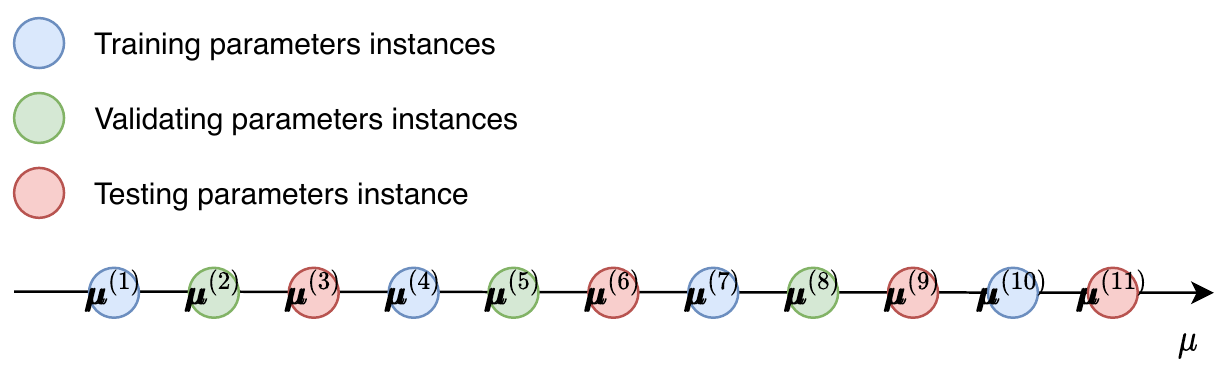} } 
    \caption{Visualizations of parameter instances sampling for the quasi-1D Euler equations.}
	\label{fig:q1e_param_grids}
\end{figure}

The varying parameter of this problem is the width of the middle cross-sectional area, which determines the geometry of the spatial domain and, thus, determines the initial condition as well as the dynamics. Analogously to the previous two benchmark problems, we select 4 training parameter instances, 3 validating parameter instances, and 3 testing parameter instances (Figure \ref{fig:q1e_param_grids}): 
\begin{align*}
\paramSpaceTrain&=\{(0.13+(0.005)k)\}, \{k\}=\{0,3,6,9)\}, \\
\paramSpaceVal&=\{(0.13+(0.005)k\}, \{k\}=\{1,4,7\}, \\
\paramSpaceTest&=\{(0.13 + (0.005)k)\}, \{k\} = \{2,5,8,10\}.
\end{align*}
Again, we set the reduced dimension as $\dofROM=5$. 

\begin{figure}[!h]
    \centering
    \begin{minipage}{.3\linewidth}
    	\centering
    	\subfloat[$\timeSymb=0.1$]    {\includegraphics[scale=.65]{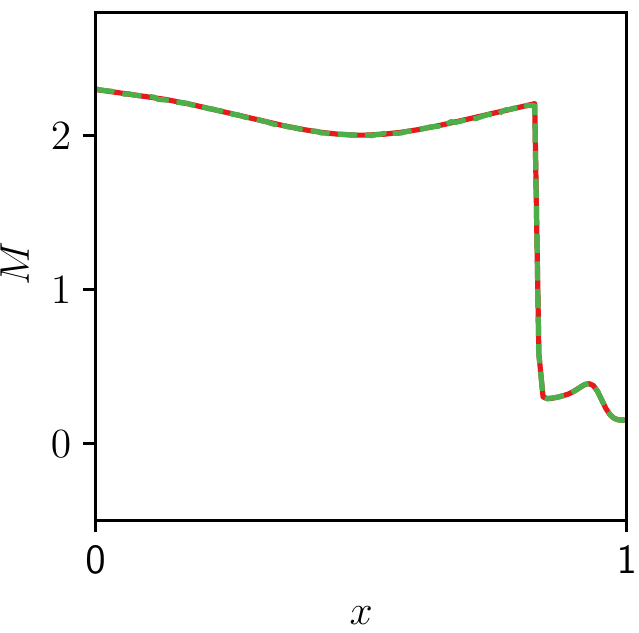}}
    \end{minipage}
    \begin{minipage}{.3\linewidth}
    	\centering
    	\subfloat[$\timeSymb=0.2$]    {\includegraphics[scale=.65]{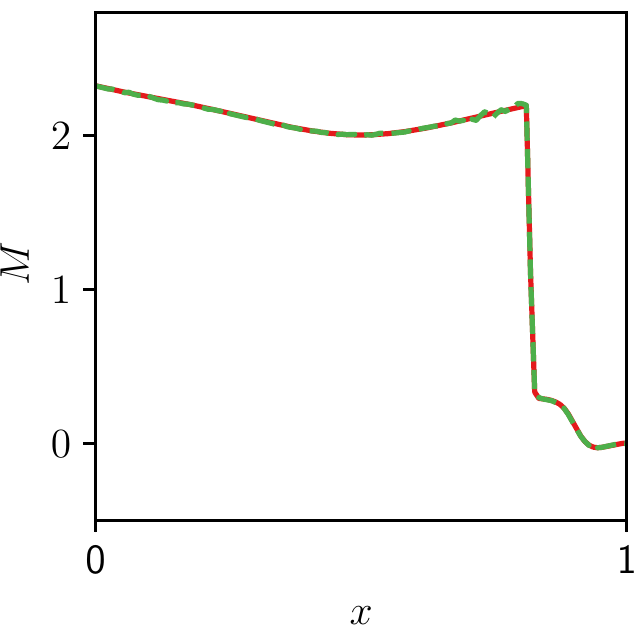}}
    \end{minipage}
    \begin{minipage}{.3\linewidth}
    	\centering
 	\subfloat[$\timeSymb=0.3$]    {\includegraphics[scale=.65]{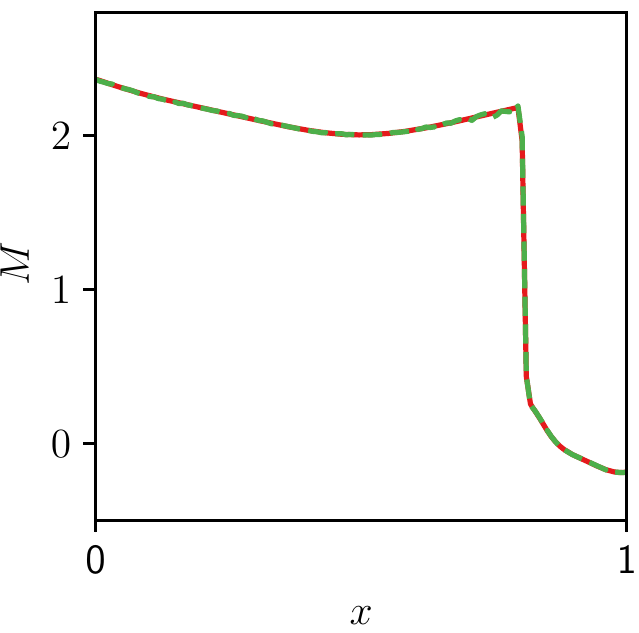}}
    \end{minipage}\\
    \begin{minipage}{.3\linewidth}
    	\centering    
	\subfloat[$\timeSymb=0.4$]    {\includegraphics[scale=.65]{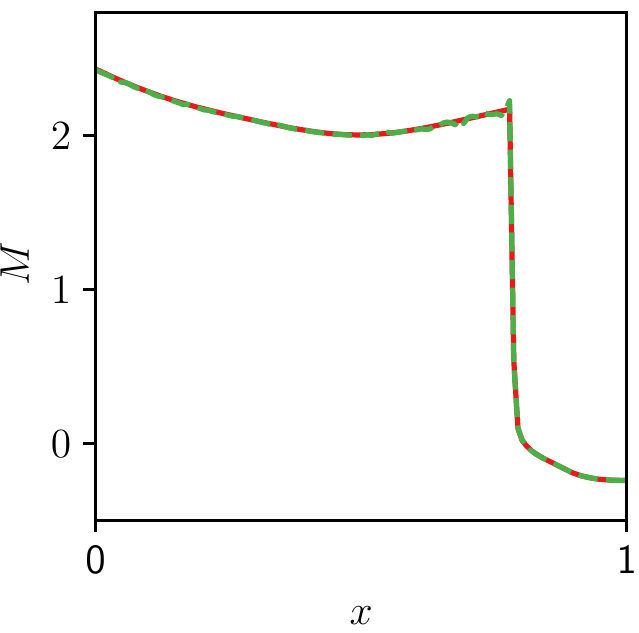}}
    \end{minipage}
    \begin{minipage}{.3\linewidth}
    	\centering    
    	\subfloat[$\timeSymb=0.5$]    {\includegraphics[scale=.65]{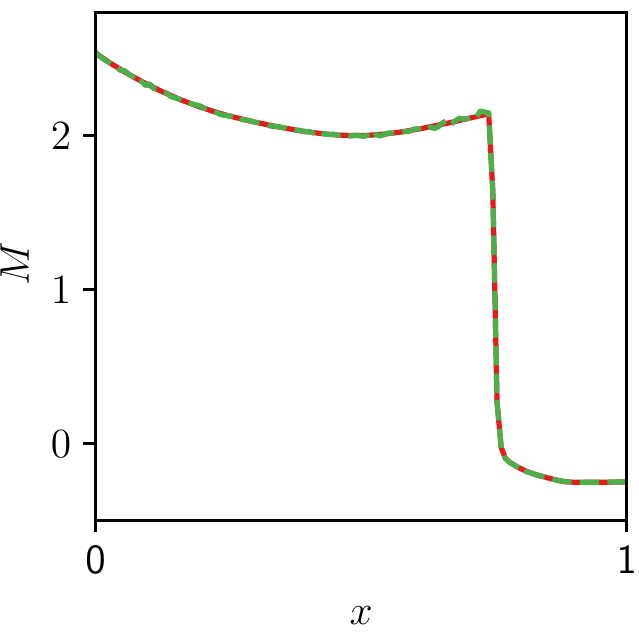}}
    \end{minipage}
    \begin{minipage}{.3\linewidth}
    	\centering
    	\subfloat[$\timeSymb=0.6$]    {\includegraphics[scale=.65]{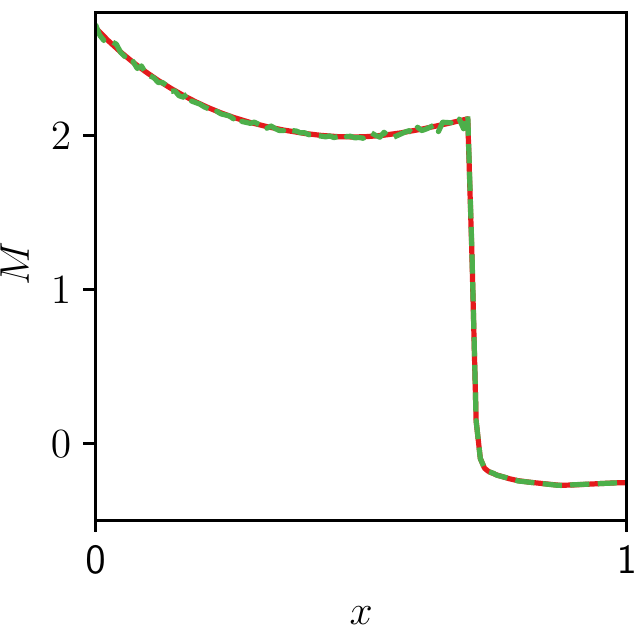}}
    \end{minipage}
    \caption{Snapshots of reference solutions (solid red lines) and approximated solutions (dashed green lines) of Mach number $\MachSymb(\positionalSymb)$ for $\paramSymb=0.15$ at $\timeSymb=\{0.1,0.2,0.3,0.4,0.5,0.6\}$. The approximated solutions are obtained by using the framework with PNODE.}
	\label{fig:euler_test_sol}
\end{figure}

We train the framework either with NODE and PNODE for learning latent dynamics and test the framework in the predictive scenario (i.e., for unseen testing parameter instances as shown in Figure \ref{fig:q1e_param_grids}) and Figure \ref{fig:euler_test_sol} depicts the solution snapshots at $\timeSymb=\{0.1,0.2,0.3,0.4,0.5,0.6\}$.
We observe that PNODE yields moderate improvements over NODE, i.e., about 20\% decrease in the relative $\ell^2$-norm of the error \eqref{eq:rel_error} for all four testing parameter instances. The improvements are not as dramatic as the ones shown in the previous two benchmark problems. We believe this is because, in this problem setting, varying the input parameter results in fairly distinct initial conditions, but does not significantly affect variations in dynamics; both the initial condition and the dynamics are parameterized by the same input parameter, the width of the middle cross-sectional area of the spatial domain.

\begin{table}[!h]
{\footnotesize
  \caption{Network architecture: kernel filter length $\kernelLength$, number of kernel filters $\nKernels$, and strides $\strides$ at each layer of (transposed) convolutional layers.}\label{tab:q1e_network_architecture}
\begin{center}
\renewcommand{\arraystretch}{1.25}
  \begin{tabular}{|c|c|} 
\multicolumn{2}{c}{Encoder}\\
\hline
\multicolumn{2}{|c|}{Conv-layer (5 layers)}\\
\hline
$\kernelLength$  &[16, \phantom{3}8, \phantom{6}4, \phantom{6}4, \phantom{64}4] \\
$\nKernels$ &[16, 32, 64, 64, 128] \\
$\strides$ &[\phantom{3}2,\phantom{3}2, \phantom{6}2, \phantom{6}2, \phantom{64}2] \\
\hline
\multicolumn{2}{|c|}{FC-layer (1 layer)}\\
\hline
\multicolumn{2}{|c|}{$\inputDim=512$, $\outputDim=\dofROM$}\\
\hline
\end{tabular}
\hspace{5mm}
\begin{tabular}{|c|c|} 
\multicolumn{2}{c}{Decoder}\\
\hline
\multicolumn{2}{|c|}{FC-layer (1 layer)}\\
\hline
\multicolumn{2}{|c|}{$\inputDim=\dofROM$, $\outputDim=512$}\\
\hline
\multicolumn{2}{|c|}{Trans-conv-layer (5 layers)}\\
\hline
$\kernelLength$  &[\phantom{4}4, \phantom{1}4, \phantom{1}4, \phantom{1}8, 16] \\
$\nKernels$ &[64, 64, 32, 16, \phantom{1}3] \\
$\strides$ &[\phantom{1}2, \phantom{1}2, \phantom{1}2, \phantom{1}2, \phantom{1}2] \\
\hline
\end{tabular}
\end{center}
}
\end{table}

Our general observation is that the benefits of using PNODE are most pronounced when the dynamics are parameterized and there is a single initial condition although PNODE outperforms NODE in all our benchmark problems. We expect to see more improvements in the approximation accuracy over NODE when the dynamics vary significantly for different input parameters, for instance, compartmental modeling (e.g., SIR, SEIR models) of infectious diseases  such as the novel corona virus (COVID-19) \cite{acemoglu2020multi,wang2020phase}, where the dynamics of transmission is greatly affected by parameters of the model, which are determined by e.g., quarantine policy, social distancing. Other potential applications of PNODEs include modeling i) the response of a quantity of interest of a parameterized partial differential equations and ii) the errors of reduced-order model of dynamical systems \cite{parish2020time}. 

Our approach shares the same limitation with other data-driven ROM approaches; it does not guarantee preservation of any important physical properties such as conservation. This is a particularly challenging issue, but there have been recent advances in deep-learning approaches for enforcing conservation laws (e.g., enforcing conservation laws in subdomains \cite{lee2019deep}, hyperbolic conservation laws \cite{raissi2019physics}, Hamiltonian mechanics \cite{greydanus2019hamiltonian,toth2019hamiltonian}, symplectic structures \cite{chen2019symplectic, jin2020sympnets}, Lagrangian mechanics \cite{cranmer2020lagrangian} and metriplectic structure \cite{hernandez2020structure}) and we believe that adapting/extending ideas of these approaches potentially mitigates the limitation of data-driven ROM approaches.

\section{Conclusions}
In this study, we proposed a parameterized extension of neural ODEs and a novel framework for reduced-order modeling of complex numerical simulations of computational physics problems. Our simple extension allows neural ODE models to learn multiple complex trajectories. This extension overcomes the main drawback of neural ODEs, namely that only a single set of dynamics are learned for the entire data distribution. We have demonstrated the effectiveness of of parameterized neural ODEs on several benchmark problems from computational fluid dynamics, and have shown that the proposed method outperforms neural ODEs. 


\section{Acknowledgments}
This paper describes objective technical results and analysis. Any subjective
views or opinions that might be expressed in the paper do not necessarily
represent the views of the U.S. Department of Energy or the United States
Government.
Sandia National Laboratories is a multimission laboratory managed and operated by National
Technology \& Engineering Solutions of Sandia, LLC, a wholly owned subsidiary
of Honeywell International Inc., for the U.S. Department of Energy's National
Nuclear Security Administration under contract DE-NA0003525.

\bibliography{ref.bib}
\bibliographystyle{siam}

\end{document}